\PassOptionsToPackage{colorlinks=true,linkcolor=blue,citecolor=blue}{hyperref}
\documentclass[prd,twocolumn,nofootinbib,superscriptaddress,letterpaper]{revtex4}

\newcommand\ForInternalReference[1]{}
\newcommand\SkipForEarlyCirculation[1]{}

\newcommand\SkipPP[1]{}
\usepackage{tabularx}
\usepackage{array}
\usepackage{graphicx}
\usepackage{dcolumn}
\usepackage{bm}
\usepackage{xspace}
\usepackage{url}
\usepackage{amsmath}
\usepackage{longtable}
\usepackage{listings}
\usepackage{multirow}
\usepackage{amssymb}
\usepackage{color}
\usepackage{tikz}
\usepackage{orcidlink}
\usepackage{cleveref}
\usepackage{xurl}
\usetikzlibrary{shapes.geometric, arrows, bayesnet}
\usepackage{hyperref}
\newcommand\optional[1]{}

\tikzstyle{startstop} = [circle, rounded corners, minimum width=1cm, minimum height=1cm,text centered, draw=black, fill=red!30]
\tikzstyle{io} = [trapezium, trapezium left angle=70, trapezium right angle=110, minimum width=2cm, minimum height=1cm, text centered, draw=black, fill=blue!30]
\tikzstyle{process} = [rectangle, minimum width=2cm, minimum height=1cm, text centered, draw=black, fill=orange!30]
\tikzstyle{decision} = [diamond, minimum width=2cm, minimum height=1cm, text centered, draw=black, fill=green!30]
\tikzstyle{arrow} = [thick,->,>=stealth]
\definecolor{amber}{rgb}{1.0, 0.75, 0.0}
\definecolor{orange}{rgb}{1.0, 0.5, 0.0}
\definecolor{amaranth}{rgb}{0.9, 0.17, 0.31}

\graphicspath{{./figures/}}

\def\ltsima{$\; \buildrel < \over \sim \;$}
\def\simlt{\lower.5ex\hbox{\ltsima}}
\def\gtsima{$\; \buildrel > \over \sim \;$}
\def\simgt{\lower.5ex\hbox{\gtsima}}

\newcommand\gwk{\textsc{GWKokab}\xspace}

\newcommand{\svec}{\lambda} 
\newcommand{\comp}{\boldsymbol{\rho}}
\newcommand{\svecz}{\boldsymbol{\lambda}} 
\newcommand{\pvecz}{\boldsymbol{\Lambda}}
\newcommand{\data}{\mathcal{D}}

\def\RIT{Center for Computational Relativity and Gravitation, Rochester Institute of Technology, Rochester, New York 14623, USA}

\begin{document}

\renewcommand{\arraystretch}{1.5}
\title{Assessing the waveform systematics from parameter estimation to population inference with eccentricity}

\author{M. Zeeshan\orcidlink{0000-0002-6494-7303}}
\email{m.zeeshan5885@gmail.com}
\affiliation{\RIT}

\author{R. O'Shaughnessy\orcidlink{0000-0001-5832-8517}}
\affiliation{\RIT}

\author{N. Malagon\orcidlink{0000-0002-5825-7795}}
\affiliation{\RIT}

\author{K. J. Wagner\orcidlink{0000-0002-7255-4251}}
\affiliation{\RIT}

\begin{abstract}


The increasing sensitivity of gravitational wave surveys is revealing a growing and increasingly diverse population of compact binary sources, including systems with possible evidence for exceptional properties such as orbital eccentricity. While masses and spins are routinely used to constrain compact-binary formation channels, eccentricity provides an additional and potentially powerful diagnostic of binary origin, particularly for dynamically assembled systems. Recent advances in eccentric waveform modeling now make it possible to search for eccentric signatures in gravitational-wave data; however, differences between waveform models can introduce systematic effects that may propagate into astrophysical population inference. In this work, we analyze 153 binary black holes, 2 binary neutron stars and 7 neutron star black hole binaries from the GWTC-4 catalog. We compare the source and population level inferences obtained with two eccentric waveform models, \textsc{SEOBNRv5EHM} and \textsc{TEOBResumS-Dali}, as well as with quasi-circular waveform analyses. We find that the two eccentric models give broadly consistent source parameter estimates for most events, but some events exhibit subtle and coherent differences. These small, systematic offsets can accumulate in hierarchical population inference, leading to differences in inferred population properties, most notably in the redshift evolution and effective spin distribution. Because coherent event level biases can grow approximately as $\sqrt{N}$ for a catalog of $N$ events, waveform systematics become increasingly important as gravitational wave catalogs expand. We also introduce a synthetic data framework that generates eccentric populations and corresponding RIFT posterior samples, enabling injection studies that test the recoverability of eccentric population properties.

\end{abstract}
\maketitle

\section{Introduction}
\label{sec:intro}

The catalog of gravitational-wave (GW) sources
\cite{2019PhRvX...9c1040A,2021PhRvX..11b1053A,2024PhRvD.109b2001A,LIGO-O3-O3bcatalog,LIGO-O4a-cbc-catalog_results} 
observed by the (LIGO-Virgo-KAGRA) LVK detectors \cite{LIGOScientific:2014pky,VIRGO:2014yos,10.1093/ptep/ptaa125} continues to grow
as detector sensitivities improve \cite{2015CQGra..32g4001L,2019NatAs...3...35K,2021PTEP.2021eA101A,2025arXiv250818081T,2020LRR....23....3A}.
The larger and more comprehensive catalog increasingly includes new discoveries, with physics or phenomena previously not
confidently apparent from earlier catalogs
\cite{LIGO-O1-BoxingDay,LIGO-GW170817-bns,LIGO-O3-GW190412,LIGO-O3-GW190814,LIGO-O3-GW190521-implications,LIGO-O4-HierarchicalPair-2025,LIGO-O4-GW231123}.
The distinctive properties of these discoveries provide clues into how compact binaries form
\cite{2009LRR....12....2S,Thorne1977,2010CQGra..27k4007M,PSconstraints3-MassDistributionMethods-NearbyUniverse,2016Natur.534..512B,AstroPaper,2017ApJ...846...82Z,2017PhRvL.119y1103V,2022PhR...955....1M,2022ApJ...940..171R,2020ApJ...903L...5R,2021ApJ...921L..43Z,2020FrASS...7...38M}.
As one example, recent observations have confidently identified multiple events each individually consistent with hierarchical compact binary
formation:  very massive black holes with large spin (GW231123\_135430 \cite{LIGO-O4-GW231123,2026arXiv260615150C}), lower-mass asymmetric binaries with large primary black holes (BH) spin \cite{LIGO-O4-HierarchicalPair-2025,2025arXiv250923897L}.
%

More broadly, however, the whole catalog combined can reveal the underlying population, enabling  sharper questions about
compact binary formation channels
\cite{LIGO-O4a-cbc-catalog_methods,2025arxiv250818083T,LIGO-O3-O3a-RP,LIGO-O3-O3bpop,gwastro-PopulationReconstruct-Parametric-Wysocki2018,gwastro-DanielW-PopsynKickPaper2017,gwastro-Davide-PopsynKickPaper2018,gwastro-PopulationReconstruct-Hierarchical-WysockiDoctor2019,popsyn-gwastro-STInterpFinal-Vera2023,gwastro-wd-DelfaveroCosmic-2024,gwastro-agndisk-GayathriPopModels2022,gwastro-agndisk-GayathriPopModels2025}.
For example, the overall distribution of binary spins and trends in primary spin versus mass may provide insight to differentiate between different formation
channels. Some of which might form binaries with preferentially aligned spins and masses generated from isolated binary stellar evolution, and others which invoke more dynamic formation scenarios including hierarchical triples, dense
clusters, or Active Galactic Nuclei (AGN) disks; see, e.g.,  \cite{2025arxiv250818083T,LIGO-O3-O3a-RP,LIGO-O3-O3bpop}
and references therein. Many recent investigations point to distinctive structures in the mass distribution \cite{2022ApJ...928..155T,2026arXiv260414290G,2026arXiv260525994G} as well as significant
changes of spin with mass, hinting
at hierarchical formation
\cite{2025arXiv250923897L,2025arXiv250717551L,2025arXiv250915646B,2025arXiv250923897L,2025arXiv250717551L,dcc-Tong-Hierarchical-2025,2025arXiv251025579T,2025PhRvL.134a1401A,2024PhRvL.133e1401L,2022ApJ...928..155T,2022PhRvD.105l3024F,2024arXiv240601679P}.
These sometimes-sharp features in turn can provide standard candles, to inform high-precision cosmology with gravitational wave
sources  \cite{2012PhRvD..85b3535T,LIGO-O3-O3bpop,2022PhRvL.129f1102E,2025arXiv250903607M,2025ApJ...985..220T,LIGO-O4b-CBC-cosmo}.

All the benefits accrued from joint population inference require population modeling whose conclusions are robust to the
underlying interpretation of each GW source.  However, GW source parameter inference is known to have often-significant
systematic uncertainties, derived principally from limited understanding of strong-field general
relativity 
\cite{gwastro-RIFT-systematics-AnjaliAasim-2020,gwastro-mergers-TousifGWTC3,gwastro-mergers-bns-systematics-Yelikar2024,gwastro-mergers-rift_asimov_O3-Fernando2024,gwastro-pe-eccentric-Malagon2026,2026arXiv260421859K} .  For example, several GW sources admit different interpretations depend
on the modeling framework used
\cite{LIGO-O3-GW190412,LIGO-O3-GW190814,LIGO-O3-GW190521-implications,2021ApJ...907L...9N,LIGO-O4-GW231123}, let alone the range of physical properties that
are a priori allowed
\cite{gwastro-nr-eccentric-190521g,gwastro-pe-eccentric-Wagner2024,gwastro-PatriciaCarlosNR-2025,gwastro-pe-eccentricGW231123-Jan2025,gwastro-pe-eccentric-Malagon2026,gwastro-hyperbolic-O3-Lange}.
Some of the broad coarse properties of the overall population may not be impacted by small biases in the interpretation
of individual events  \cite{2025PhRvX..15c1036D,2024PhRvD.110d3520K,2026arXiv260528716P,2026arXiv260516510D}.
However, subtle features and subpopulations \cite{2026arXiv260602318P,2026arXiv260600234P,2025arXiv250909123A,2025PhRvD.112b3531A,2023MNRAS.523.4539K} whose properties are localized to regions of substantial systematics
inevitably will be impacted by systematics.

In this work, we identify a new challenge for source population inference: the subtle, consistent accumulation of small
biases in source interpretation between  different waveform models.  Specifically, we identify a small consistent bias in
the recovered distances as inferred using two state-of-the-art GW waveform models, when applied to the GWTC-4 source
sample. As a result, we demonstrate that the recovered redshift distribution is not a robust observable at the level of
accuracy these two models seem to allow.
We also identify other less-significant secular biases between these two waveforms, which in a larger sample may also
lead to consistent differences accumulated across the whole population


This paper is organized as follows.
In Section~\ref{sec:methods}, we review our method for interpreting the GW census, the specific population models used
for real and synthetic data,
and the GW source parameter inferences used as inputs.
In Section~\ref{sec:results}, we demonstrate the systematic biases between source inferences using simple frequentest diagnostics, then demonstrate how these inferences propagate into our source population inference with concrete
end-to-end calculations.
We also report on two other calculations using our population inference framework applied to eccentric source parameter
inference.  First, motivated by strong indications of eccentricity in GW200105\_162426,  we characterize the population of NS-BH
binaries.  Second, we validate our framework using end-to-end source and population inference using synthetic data.
We conclude in Section~\ref{sec:conclude}.

\section{Methods}
\label{sec:methods}
\subsection{Review of hierarchical Bayesian inference (HBI)}\label{subsec:hbi_review}

To infer the population properties, we adopt the formalism introduced in previous works, referred to as Bayesian parametric models (BPM), implemented in the population inference engine called \gwk \cite{gwkokab2024github,2026PhRvD.113j3003Q}. Given the likelihood $\ell(\svecz)$ of individual sources and associated reference prior $\pi(\svecz)$, we proceed with a hierarchical Bayesian framework given in Equation~\ref{eq:Bayes_pop} to infer the posterior distribution $p(\pvecz|\data)$ of the BBH population,

\begin{align}
    \label{eq:Bayes_pop}
    \!p\left(\pvecz | \data \right)
     & = \frac{
        \pi(\pvecz)\,
        p(\data | \pvecz)
    }{
        p\left( \data \right)
    },
\end{align}
where $\data=\{{d_j}\}_{j=1}^N$ is the dataset and $d_j$ shows an individual event and $N$ is the total number of events, $p(\pvecz | \data)$ is the posterior distribution of $\pvecz$ given $\data$, $\pi(\pvecz)$ is the population prior on $\pvecz$.  The term $p(\data)$, known as Bayesian evidence, serves as a normalization constant and often omitted in sampling-based inference.
Therefore, in practice, we will use the likelihood function $\mathcal{L}(\pvecz)\equiv p(\data | \pvecz)$ to compute the posterior distribution $p(\pvecz | \data) \propto \mathcal{L}(\pvecz)~\pi(\pvecz)$.

To conduct our analysis we have used the inhomogeneous Poisson process \cite{2019MNRAS.486.1086M,2004AIPC..735..195L,PhysRevD.91.023005}

\begin{equation}
    \label{eq:likelihood}
    \mathcal{L}(\pvecz) \propto
    e^{-\mu{(\pvecz)}}
    \prod_{j=1}^N
    \int\ell_j(\svecz) \cdot \comp(\svecz\mid\pvecz)
     \sqrt{ g_{\svecz}}
    d \svecz,
\end{equation}
where exponent $\mu{(\pvecz)}$ is the total expected number of detections under the
given population parametrization $\pvecz$, the complete expression is given in Equation~\ref{eq:mu}. $g_{\svecz}$ is the determinant of the
metric over those coordinates, and  $\comp (\svec | \pvecz)$ is the merger rate density in source frame of reference. For source-parameters, we adopt a usual uniform metric over all intrinsic and extrinsic parameters, such that $\sqrt{g_{\svecz}}d\svecz = T_{\mathrm{obs}} \times dz (1+z)^{-1} (dV_c/dz) \times dm_1 dm_2 \times $  appropriate factors for eccentricity and spin which depend on the coordinate representation adopted for them. The term $\ell_j(\svecz)$ is
the likelihood of individual events and defined as follows,

\begin{equation}
\ell_j(\svecz) \equiv p(d_j|\svecz) \propto \frac{ p(\svecz|d_j)}{\pi(\svecz)}.\label{eq:indi_likelihood}
\end{equation}

For the events till O3b observing runs, the euclidean reference prior $\pi(\svecz)$ is given as follows,
\begin{equation}
    \pi(\svecz)=D_L^2(z)\frac{\partial D_L(z)}{\partial z}\times(1+z)^2,
    \label{eq:ref_prior}
\end{equation}
and for events in O4a observing run, the comoving reference prior $\pi(\svecz)$ is given as follows,
\begin{equation}
    \pi(\svecz)=\frac{1}{V_0}\frac{dV_c}{dz}\frac{1}{1+z}\times(1+z)^2,
    \label{eq:ref_prior_comoving}
\end{equation}
where the factor $(1+z)^2$ converts detector-frame to source-frame masses (primary and secondary), $D_L^2(z)\partial D_L/\partial z$ corresponds to the luminosity-distance prior, and $V_0=\int_{0}^{z_{\rm max}} \frac{dV_c}{dz} \frac{1}{1+z} dz$ is normalization constant, further details are given in \cite{2021arXiv210409508C}. In this work, we model only the population distribution of $\chi_{\rm eff}$ and therefore marginalize over all remaining spin degrees of freedom, neglecting the precessing spin parameter $\chi_p$. The reference prior being used on $\chi_{\rm eff}$ is given in Equation 10 of \cite{2021arXiv210409508C}. All integrals appearing explicitly or implicitly in the expressions are computed via Monte Carlo integration, as described in \cite{2026PhRvD.113j3003Q}. Posteriors on hyperparameters $\pvecz$ are also filtered with the variance cut threshold of 1; See Eqs. 9--11 of \cite{2025PhRvD.111f3043H} for the variance calculation of the population likelihood.

The expected number of GW detections can be formulated as an integral over the intrinsic source-parameter space $\svec$ and redshift $z$ modulated by an appropriate selection (weighting) function. The total expected number of detections summing over all populations is given by

\begin{align}
    \label{eq:mu}
    \mu(\pvecz)= \int P_{\mathrm{det}} (\svec;z)\cdot \comp(\svecz\mid\pvecz)\sqrt{ g_{\svecz}}
    d \svecz.
\end{align}

Here $P_{\mathrm{det}}(\svec;z)$ is the detection probability for a source with intrinsic parameters $\svec$ at redshift $z$. For the synthetic population, we have generated the semi-analytical VT (not calibrated on real sensitivity injections) based on the method described in the Section II C \cite{2026PhRvD.113j3003Q}. Specifically, to include the eccentricity in selection effects, we use the \textsc{TaylorF2Ecc} waveform model restricting the eccentricity up to 0.3 and used \textsc{SimNoisePSDaLIGO175MpcT1800545} model as power spectral density. For real data analyses, we have used the GWTC-4 sensitivity injections as used in previous studies by ignoring the eccentricity from selection effects, details are in Section III B \cite{2026arXiv260211030Z}.

\subsection{Population model for BBH}
\label{subsec:bbh_population_model}
For the BBH analysis, we adopt the default parametric population model used in the GWTC-4 population study \cite{2025arxiv250818083T}, augmented with an explicit eccentricity distribution.  The mass distribution is modeled with the \textsc{Broken Power Law + 2 Peaks} prescription, defined in Eq.~B13 of Ref.~\cite{2025arxiv250818083T}, with the corresponding hyperparameter priors listed in their Table~6.  The merger-rate evolution with redshift is described by the \textsc{Power Law} model of Eq.~B25, using the priors reported in Table~7 of the same reference.

For spins, we model only the effective inspiral spin $\chi_{\rm eff}$ and adopt the \textsc{Skew-Normal Effective Spin} model, defined in Eq.~B37 of Ref.~\cite{2025ApJ...990..147B,2025arxiv250818083T}, with priors from their Table~9.  This choice is appropriate for our analysis because the skew-normal model does not require an explicit correlation between $\chi_{\rm eff}$ and the precessing spin parameter $\chi_p$.  We can therefore infer the population distribution of $\chi_{\rm eff}$ after marginalizing over the remaining spin degrees of freedom; see Sec.~6.3.2 of Ref.~\cite{2025arxiv250818083T} and references therein for further discussion.

Compared to our previous eccentric BBH population analysis \cite{2026arXiv260211030Z}, we replace the mixture eccentricity prescription with a simpler power-law eccentricity distribution for the real-data analysis.  Specifically, we take
\begin{equation}
    p(e\mid \alpha_e) \propto e^{\alpha_e},
    \qquad
    10^{-6} \leq e \leq 0.5,
\end{equation}
with a uniform prior $\alpha_e\in[-5,5]$.  This single-parameter extension provides a compact way to test whether the observed BBH population prefers enhanced support at larger or smaller eccentricities while keeping the baseline GWTC-4 mass, spin, and redshift assumptions fixed.

\subsection{Population model for BNS and NSBH}
\label{subsec:bns_nsbh_population_model}

We extend the phenomenological population model used in the GWTC-3 population analysis for a joint inference of BNS and NSBH sources.  The present sample contains only a small number of neutron-star binaries, so we adopt a deliberately simple model that shares information across BNS and NSBH systems while still allowing the black-hole component in NSBH binaries to have a distinct mass and spin distribution.  This choice reduces the number of hyperparameters and avoids over-interpreting features that cannot be robustly constrained by the current catalog.

For the mass distribution, all neutron-star components are described by a common truncated Gaussian distribution.  This common neutron-star component is used for both masses in BNS systems and for the neutron-star mass in NSBH systems.  The black-hole mass in NSBH systems is modeled with a separate truncated Gaussian distribution.  Therefore, the model assumes that neutron stars in BNS and NSBH binaries are drawn from the same underlying neutron-star mass distribution, while the black holes in NSBH binaries are allowed to occupy a different mass scale.

We use an analogous construction for the spin magnitude distribution. All neutron star spin magnitudes share a common population distribution, whereas the black-hole spin magnitude in NSBH systems is assigned its own independent distribution. This parameterization captures the expectation that neutron stars and black holes may have different spin populations, but avoids introducing separate spin distributions for the two neutron stars in BNS systems or for neutron stars in BNS versus NSBH binaries.

For eccentricity and redshift evolution, we assume a common distribution for both BNS and NSBH sources.  Specifically, we use the same power-law eccentricity model as in the BBH analysis, truncated over the same eccentricity interval, and a common power-law redshift evolution for the merger-rate density.  With these assumptions, the joint BNS and NSBH model can be interpreted as a minimal multi-source population model: it distinguishes neutron-star and black-hole component properties where the data are most physically expected to differ, while tying together weakly constrained degrees of freedom across source classes. The complete description of parameters and their prior ranges are given in Table~\ref{tab:bns_nsbh_priors}

\begin{table*}[t]
\caption{\label{tab:bns_nsbh_priors}
Model parameters and their prior ranges for BNS and NSBH joint analysis.
}
\begin{ruledtabular}
\begin{tabular}{lll}
Parameter & Prior range & Description \\
\hline
$\log R_{\rm BNS}, \log R_{\rm NSBH}$ & $\mathcal{U}(0,10)$ & Log merger rate for BNS and NSBH \\
$ \mathrm{High}_m^{\rm NS} $ & $\mathcal{U}(2,3)\,M_\odot$ & Upper truncation mass \\
$ \mu_{m}^{\rm NS} $ & $\mathcal{U}(1,3)\,M_\odot$ & Mean of neutron-star mass distribution \\
$ \sigma_{m}^{\rm NS} $ & $\mathcal{U}(0.05,3)\,M_\odot$ & Width of neutron-star mass distribution \\
$ \mu_{m}^{\rm BH} $ & $\mathcal{U}(3,50)\,M_\odot$ & Mean of black-hole mass in NSBH \\
$ \sigma_{m}^{\rm BH} $ & $\mathcal{U}(0.4,20)\,M_\odot$ & Width of black-hole mass in NSBH \\
$ \mu_{\chi}^{\rm NS} $ & $\mathcal{U}(0,0.05)$ & Mean of neutron-star spin magnitude \\
$ \sigma_{\chi}^{\rm NS} $ & $\mathcal{U}(0.001,0.05)$ & Width of neutron-star spin distribution \\
$ \mu_{\chi}^{\rm BH} $ & $\mathcal{U}(0,0.1)$ & Mean black-hole spin magnitude in NSBH \\
$ \sigma_{\chi}^{\rm BH} $ & $\mathcal{U}(0.001,0.1)$ & Width of black-hole spin distribution in NSBH \\
$ \alpha_e $ & $\mathcal{U}(-3,3)$ & Eccentricity power-law index for common BNS and NSBH \\
$ \kappa $ & $\mathcal{U}(-5,5)$ & Redshift evolution index for common BNS and NSBH \\
\end{tabular}
\end{ruledtabular}
\end{table*}

\subsection{Population model for synthetic events}\label{subsec:pop_model_syn}

\begin{figure}
    \centering
    \includegraphics[width=0.45\textwidth]{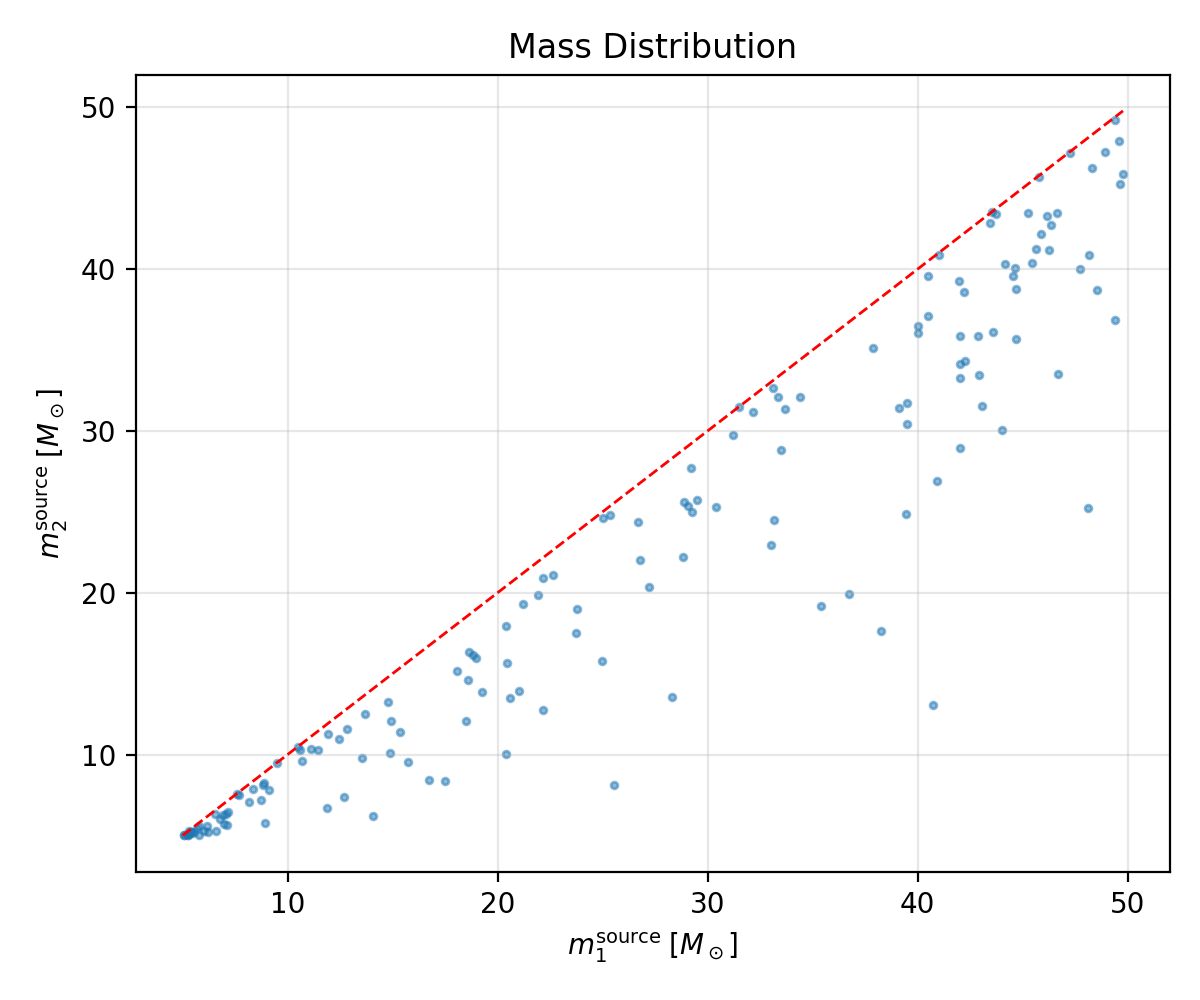} 
    \caption{\textbf{Synthetic Population:} This shows the synthetic population generated with the population model given in \eqref{eqn:syn_pop_model}.}
    \label{fig:mass_distribution_synthetic}
\end{figure}

There are various weak and pure phenomenological population models proposed in previous studies 
\cite{2016PRXAbbot_BBH_model,2017FishBach_BBH_model,2018talbot_bbh_model,2024phrvd.110f3009z}. In our analysis, we used the pure truncated power law defined for primary mass and mass-ratio, truncated normal for spin magnitude and powerlaw for eccentricity.
We also assume that non-zero probability density only exists for $m_{min}\leq m_2 \leq m_1 \leq m_{max}$. The generalized form of the population model with parameters $\Lambda \equiv (\mathcal{R}, \alpha, \beta, m_{min}, m_{max}, \mu_\chi, \sigma_\chi, \alpha_e)$ is given as follows,

\begin{align}
p(m_1, q|\alpha,\beta,m_{\min},m_{\max})
=
C_m\,
m_1^{-\alpha} q^{\beta},
\end{align}

where $C_m$ is the normalization constant, $\alpha$ and $\beta$ are the power law index, $\mathcal{R}$ is the merger rate, $m_{min}, m_{max}$ are the minimum 
and maximum masses of the binary components in the population. The spin magnitudes are modeled with a truncated normal distribution on the physical interval $[0,1]$, assuming identical spins for primary and secondary BBH. The eccentricity distribution is modeled with half-normal distribution as described in Zeeshan et. al \cite{2024phrvd.110f3009z}. The redshift distribution is modeled with power-law same for BBH population on real data.Assuming that masses, spin magnitudes, eccentricity, and redshift are independent population properties, the full population model for synthetic data is as follows and synthetic population generated with this model is shown in Figure~\ref{fig:mass_distribution_synthetic}. The complete list of parameter priors is give in Table~\ref{tab:syn_pop_priors}

\begin{align}\label{eqn:syn_pop_model}
p(m_1,q,\chi,e,z \mid \Lambda)=C \times p(m_1,q)p(\chi)p(e)p(z).
\end{align}

\begin{table*}[t]
\caption{\label{tab:syn_pop_priors}
Fiducial model parameters and their prior ranges for the synthetic population. The true values are also shown here to generate the  synthetic catalog.}
\begin{ruledtabular}
\begin{tabular}{llll}
Parameter & True value & Prior range & Description \\
\hline
$\ln R$ & $2.5$ & $\mathcal{U}(-6,6)$ & Log merger rate of the synthetic BBH population \\
$\alpha_m$ & $1.0$ & $\mathcal{U}(-4,6)$ & Primary-mass power-law index \\
$\mu_{\chi}$ & $0.0$ & $\mathcal{U}(-1,1)$ & Location parameter of the aligned-spin distribution \\
$\sigma_{\chi}$ & $0.2$ & $\mathcal{U}(0,1)$ & Width of the aligned-spin distribution \\
$\sigma_e$ & $0.15$ & $\mathcal{U}(0,0.4)$ & Width of the eccentricity distribution \\
$\kappa$ & $2.7$ & $\mathcal{U}(-5,5)$ & Redshift-evolution power-law index \\
\end{tabular}
\end{ruledtabular}
\end{table*}

\subsection{Real GW observations of compact binaries}
\label{subsec:real_gw_observations}
A previous study \cite{2026arXiv260512818M} analyzed 
162 events from GWTC-2 \cite{LIGO-O3-O3a-catalog}, GWTC-2.1 \cite{LIGO-O3-O3a_final-catalog},
GWTC-3 \cite{LIGO-O3-O3b-catalog}, and GWTC-4 \cite{LIGO-O4a-cbc-catalog_results}.
This study adopted the same sample used for global population inference for GWTC-3 and GWTC-4: a false alarm rate (FAR) of less than one per year.
In this work, we employ their analysis results and inherit all of their parameter inference settings. Parameter inferences were performed using \textsc{SEOBNRv5EHM}
\cite{2025PhRvD.112d4038G} and \textsc{TEOBResumS-Dali}
\cite{
2020PhRvD.101j1501C,%
2024PhRvD.110h4001N,%
2025PhRvD.111f4050N}. In both cases, this investigation employed all available higher-order modes supplied by these models. In this work, for population inference, we use the same set of events as used in the GWTC-4 population paper \cite{2025arxiv250818083T} with the same selection criteria, which includes 153 BBH. The BNS and NSBH binaries are listed in the Table~\ref{tab:bns_nsbh_events} which we have used for population inference.

\begin{table}[t]
\caption{\label{tab:bns_nsbh_events}
List of BNS and NSBH events used for population inference, total are 2 BNS and 7 NSBH.
}
\begin{ruledtabular}
\begin{tabular}{ll}
Event ID & Event category \\
\hline
\texttt{GW170817\_124104} & BNS \\
\texttt{GW190425\_081805} & BNS \\
\texttt{GW190426\_152155} & NSBH \\
\texttt{GW190814\_211039} & NSBH \\
\texttt{GW190917\_114630} & NSBH \\
\texttt{GW200105\_162426} & NSBH \\
\texttt{GW200115\_042309} & NSBH \\
\texttt{GW230518\_125908} & NSBH \\
\texttt{GW230529\_181500} & NSBH \\
\end{tabular}
\end{ruledtabular}
\end{table}

\subsection{Parameter inference with RIFT }
\label{subsec:rift_parameter_inference}

Coalescing compact binaries, such as binary black holes, with quasi-circular orbits can be characterized by 15 parameters with an additional parameter used to describe tidal deformability for binary neutron stars (BNSs) or neutron star-black holes (NSBHs). The eight intrinsic parameters ($\lambda$) refer to the physical properties of the individual components within each binary including the component masses, spin magnitudes, spin tilts, and azimuthal angles. The seven extrinsic parameters ($\theta$) refer to the spacetime location and orientation of the binary system: right ascension, declination, luminosity distance, inclination, polarization, orbital phase, and coalescence time. Two additional intrinsic parameters, eccentricity and mean anomaly, are inferred to describe an eccentric aligned-spin binary system.

\begin{figure}
    \centering
    \includegraphics[width=0.45\textwidth]{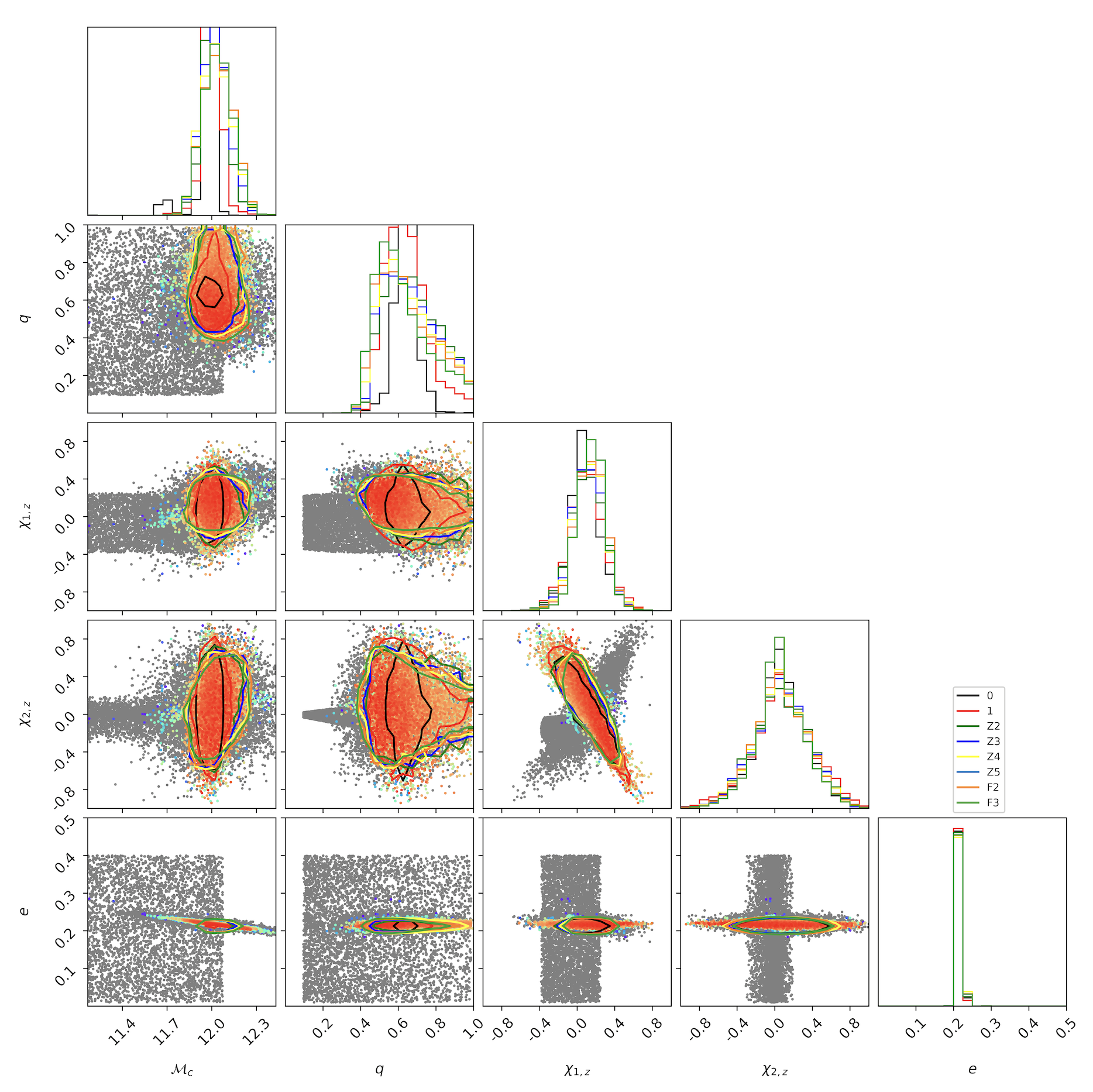} 
    \caption{This figure shows the parameter inference for a synthetic event, producing realistic posterior samples using RIFT. This method of producing posterior samples is used for all synthetic events and is the same method used for parameter estimation of real events.}
    \label{fig:rift_posterior}
\end{figure}

RIFT comprises a two-step iterative process that compares gravitational-wave observations $d$ to predicted gravitational-wave signals $h(\bm{\lambda}, \bm\theta)$. In the first step, for a large number of $\bm{\lambda}$ values, RIFT computes a marginal likelihood
\begin{equation}
 {\cal L}{({\bm \lambda})}\equiv\int  {\cal L}_{\rm full}(\bm{\lambda},\bm\theta )p(\bm\theta )d\bm\theta
\end{equation}
where ${\cal L}_{\rm full}(\bm{\lambda},\bm{\theta})$ is the likelihood of the GW signal in the multi-detector
network \cite{gwastro-PENR-RIFT,gwastro-RIFT-Update,gwastro-RIFT_FinerNet}. 
During the second step, RIFT approximates ${\cal L}(\bm{\lambda})$ based on the accumulated marginal likelihood evaluations $(\bm{\lambda},{\cal L})$. This approximation is used to construct the detector-frame posterior distribution
\begin{equation}
\label{eq:post}
p_{\rm post}(\bm{\lambda})=\frac{{\cal L}(\bm{\lambda} )p(\bm{\lambda})}{\int d\bm{\lambda} {\cal L}(\bm{\lambda} ) p(\bm{\lambda} )}.
\end{equation}
where the prior $p(\bm{\lambda})$ denotes the prior distribution on the intrinsic parameters.
After intrinsic inference is complete, extrinsic parameters $\theta$ are generated for each intrinsic sample via Monte Carlo. The resulting set of posterior samples are shown in Figure~\ref{fig:rift_posterior} for a single event selected from the Figure~\ref{fig:mass_distribution_synthetic}.

\section{Results}
\label{sec:results}

\subsection{Identifying subtle differences in source parameter inference over a population}
\label{subsec:event_level_systematics}
Parameter inferences produced by the two different waveform models have few events with immediately-apparent large-scale differences in selected parameters (as previously noted in Malagon et al \cite{2026arXiv260512818M})
and subtle, consistent, systematic differences are only apparent when assessing at large catalog of many events.  On an event-by-event basis, the differences between these two approaches can be very small; however, as we will demonstrate later, these small consistent differences accumulate, producing substantial changes on some key large-scale observables.

\begin{figure*}
    \centering
    \includegraphics[width=0.97\textwidth]{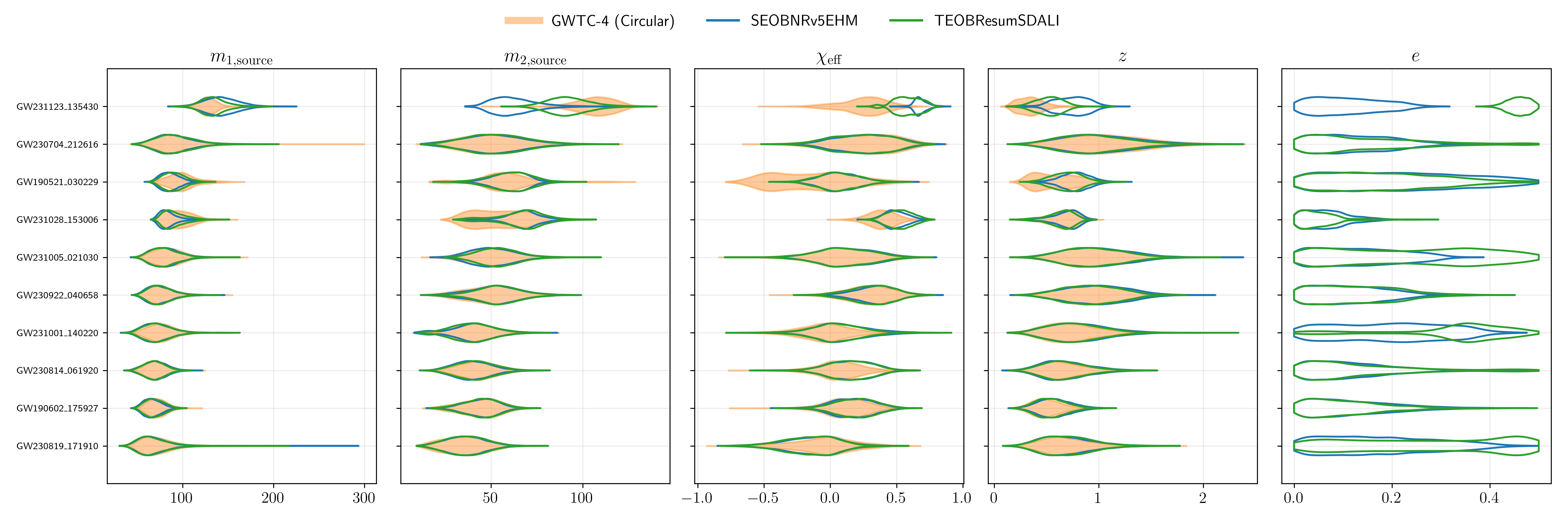}
    \caption{This shows the comparison of GWTC-4 studies assuming quasi-circular binaries vs RIFT PE using \textsc{SEOBNRv5EHM} and \textsc{TEOBResumS-Dali}. The events are sorted based on the primary mass. }
    \label{fig:overlay_pes}
\end{figure*}

Figure~\ref{fig:overlay_pes} illustrates how a few events exhibit substantial differences in selected parameters, both
between our two models and relative to previously-published quasi-circular analysis.   Even among these most extreme
single-event differences by JS divergence, however, only a handful exhibit notable large-scale differences in their
mass, distance, or spin posteriors.  We highlight these events for reference; later, we will demonstrate that removing
some or all of these events does not appreciably impact our conclusions derived from joint population inference.

\begin{figure*}
    \centering
    \includegraphics[width=0.49\textwidth]{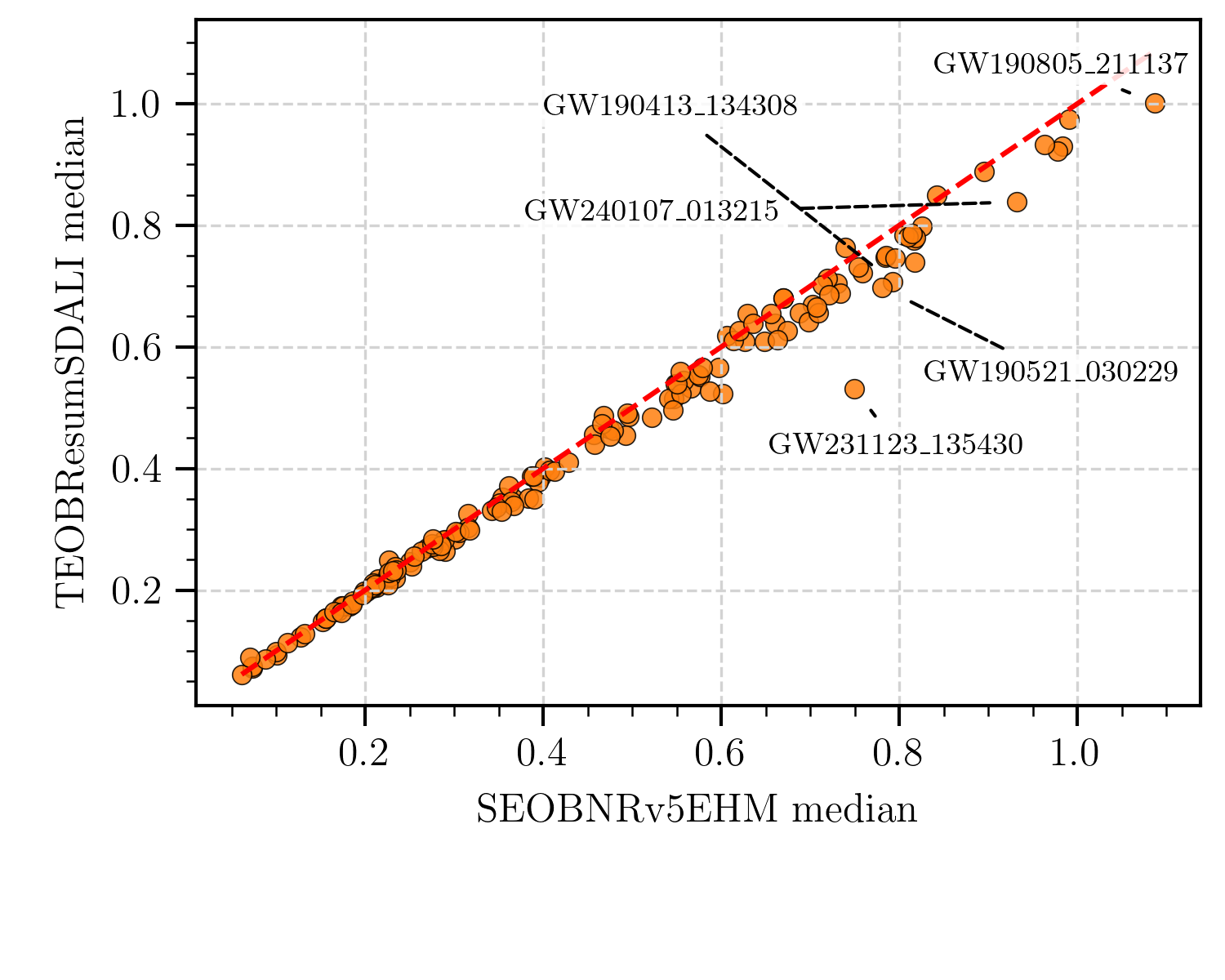}
    \includegraphics[width=0.49\textwidth]{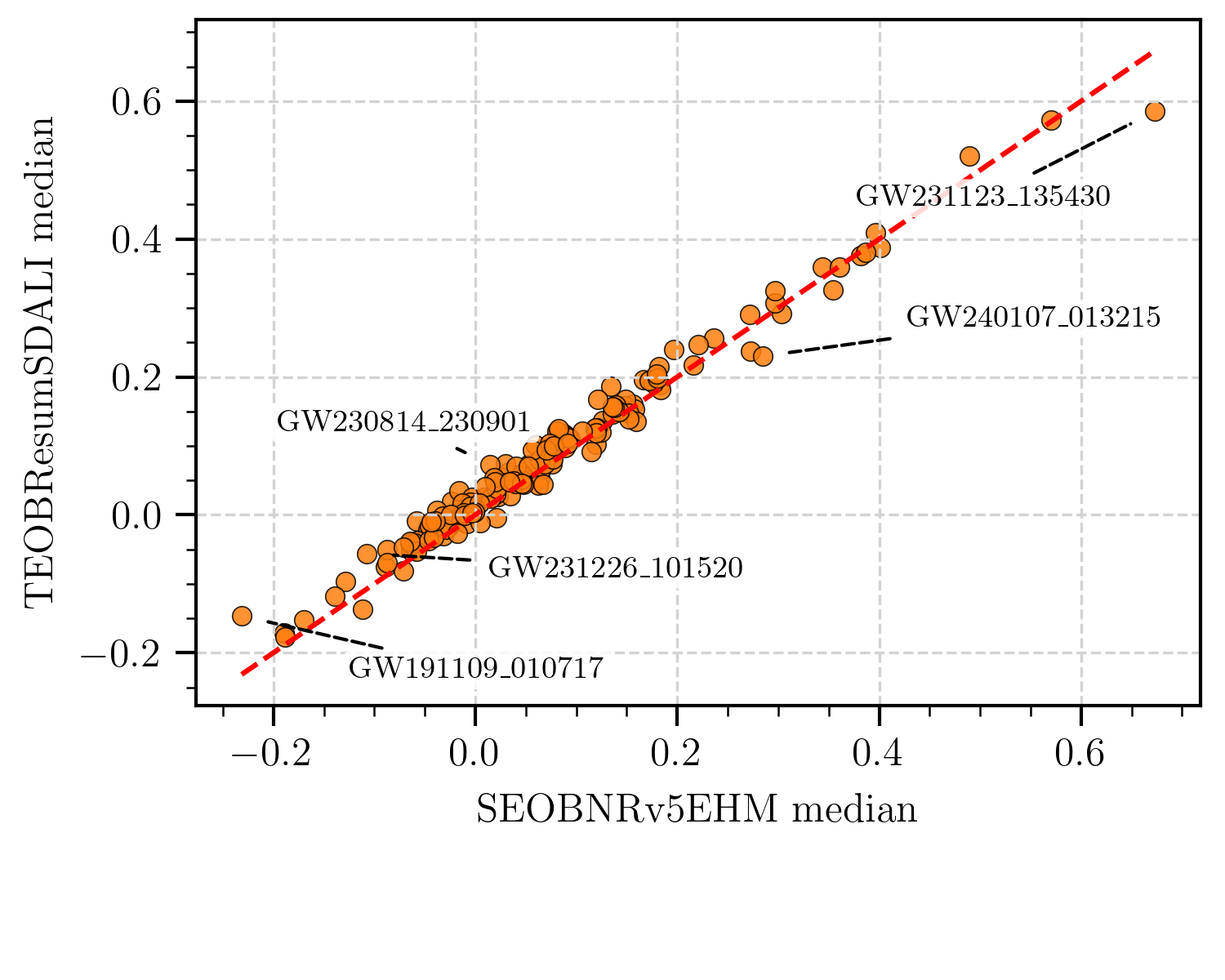}
    \includegraphics[width=0.49\textwidth]{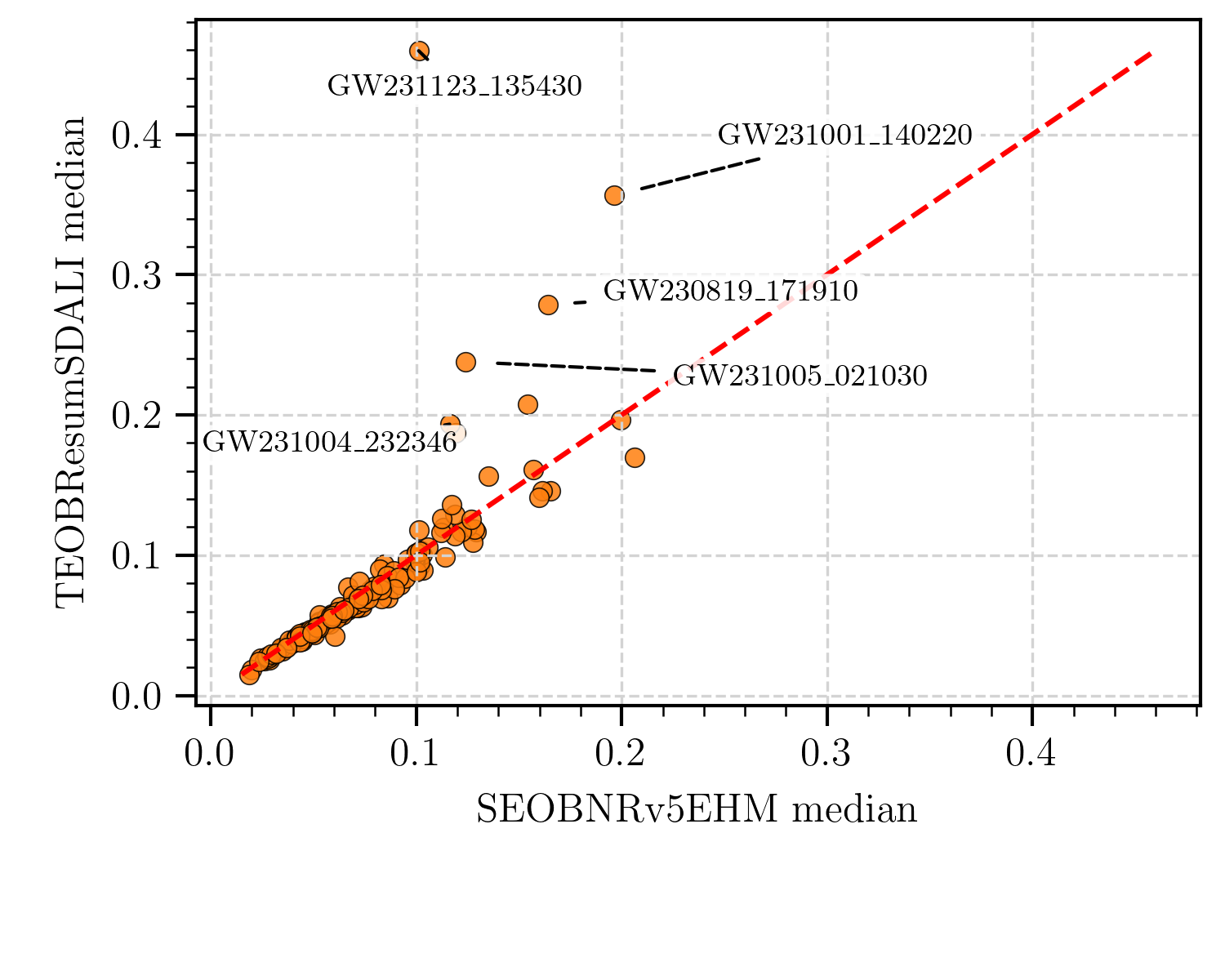}
    \includegraphics[width=0.49\textwidth]{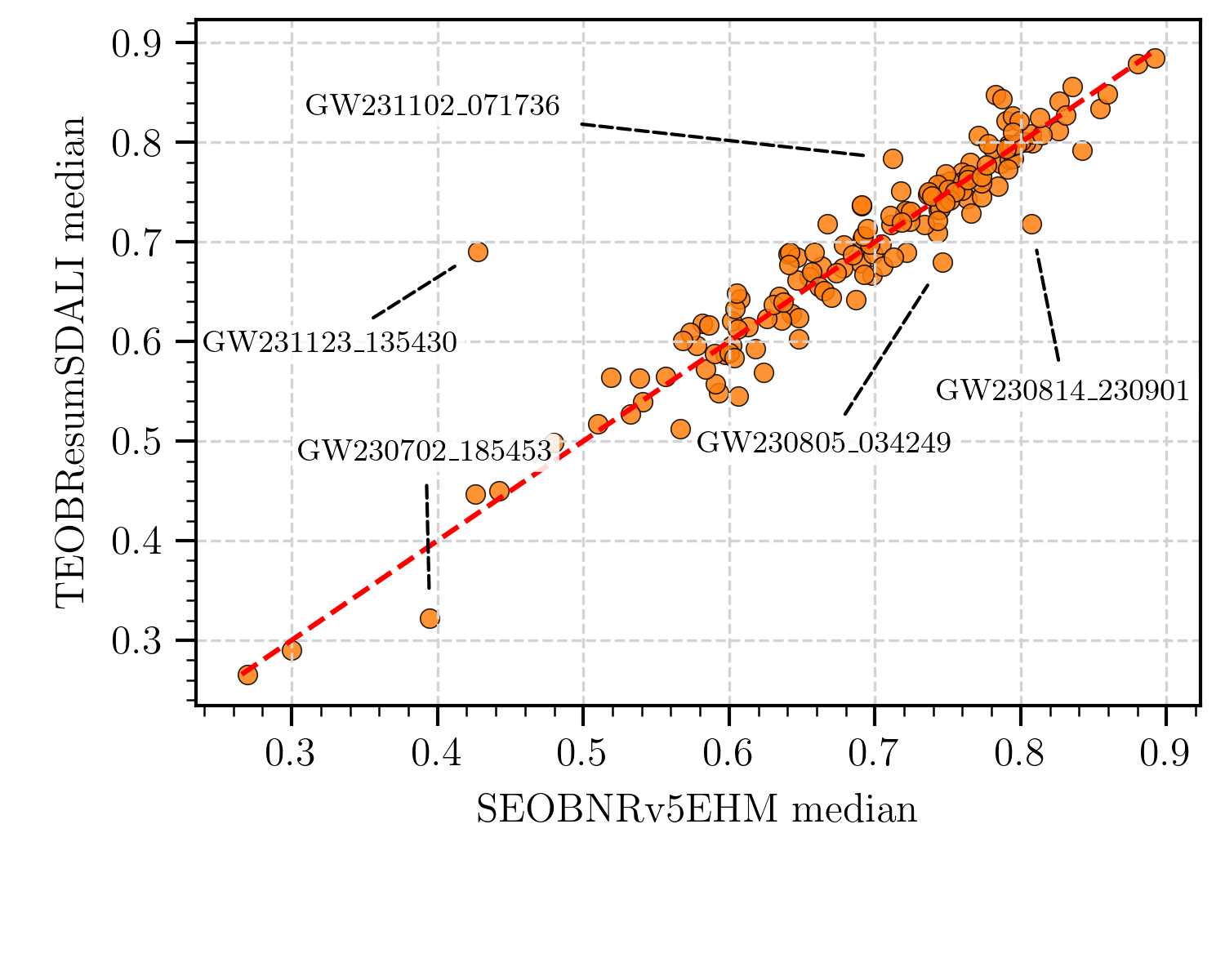}
    \caption{This figure shows the median comparison for each event, and red dashed diagonal lines shows where both models infer same median, the events with top five median offset are also labeled for identification. The upper left panel shows the redshift, upper right panel shows the effective spin, lower left panel shows the eccentricity and lower right panel shows the mass ratio. The redshift and effective-spin distributions show stronger systematic effects than eccentricity and mass ratio.}
    \label{fig:median_comparison}
\end{figure*}

Figure~\ref{fig:median_comparison} consist of four panels, each showing a scatter-plot comparing the median value of a different parameter inferred with \textsc{SEOBNRv5EHM} (horizontal axis) and \textsc{TEOBResumS-Dali} (vertical axis).  Each of these four panels
illustrates how we can assess subtle systematic bias on a population scale, even though on an individual-event basis
differences are almost always small.  For redshift, \textsc{SEOBNRv5EHM} inferences for the most distant binaries are consistently farther
than those derived with \textsc{TEOBResumS-Dali}.  For effective inspiral spin, \textsc{TEOBResumS-Dali} consistently recovers slightly more
positive values for the events with the smallest $\chi_{\rm eff}$.   For eccentricity, as noted in the previous work,
the two models are very consistent for most events, but for events with substantial eccentricity \textsc{TEOBResumS-Dali} is
consistently larger. For binary mass ratio, however, the two models are largely
consistent.

\begin{figure*}
    \centering
    \includegraphics[width=0.49\textwidth]{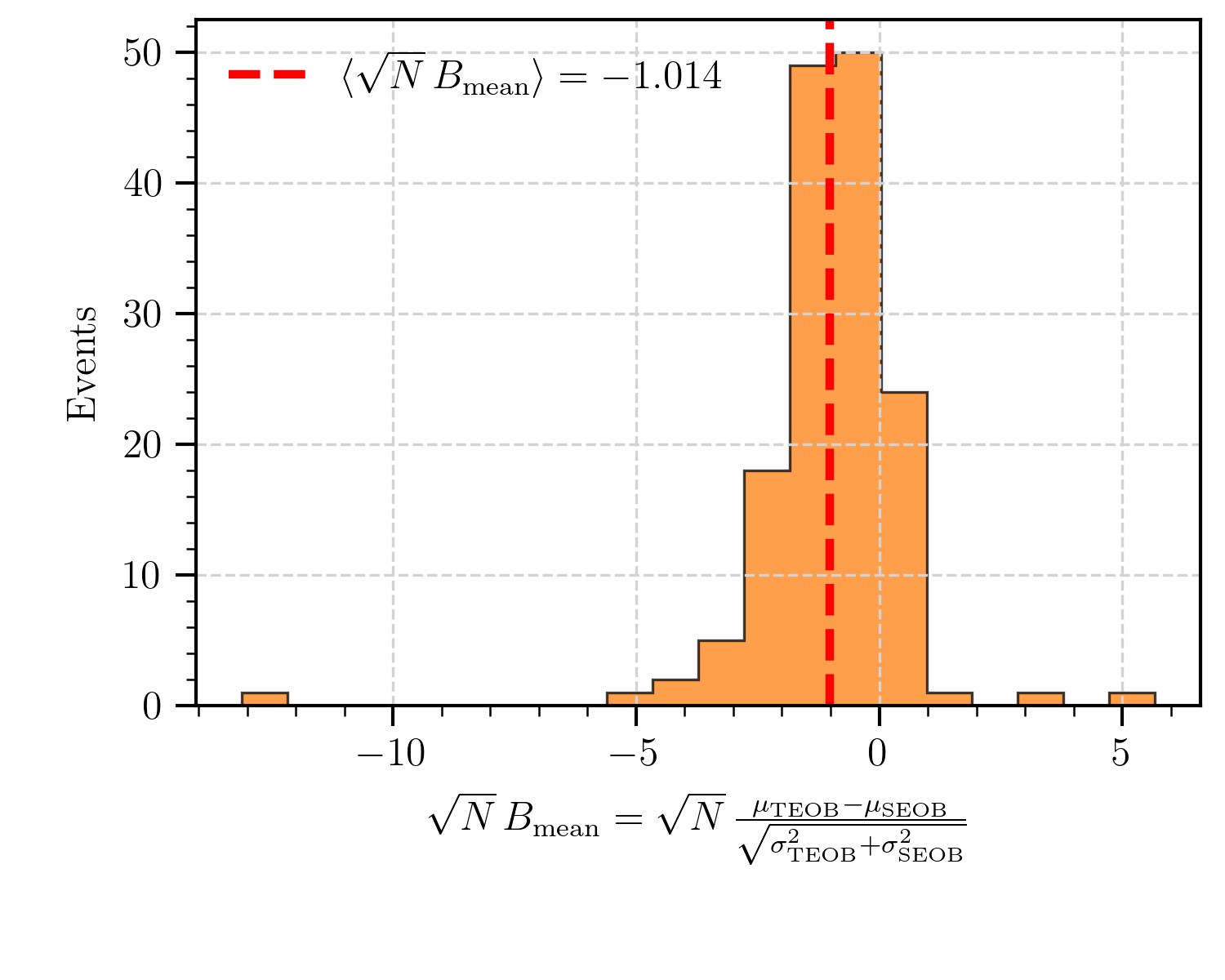}
    \includegraphics[width=0.49\textwidth]{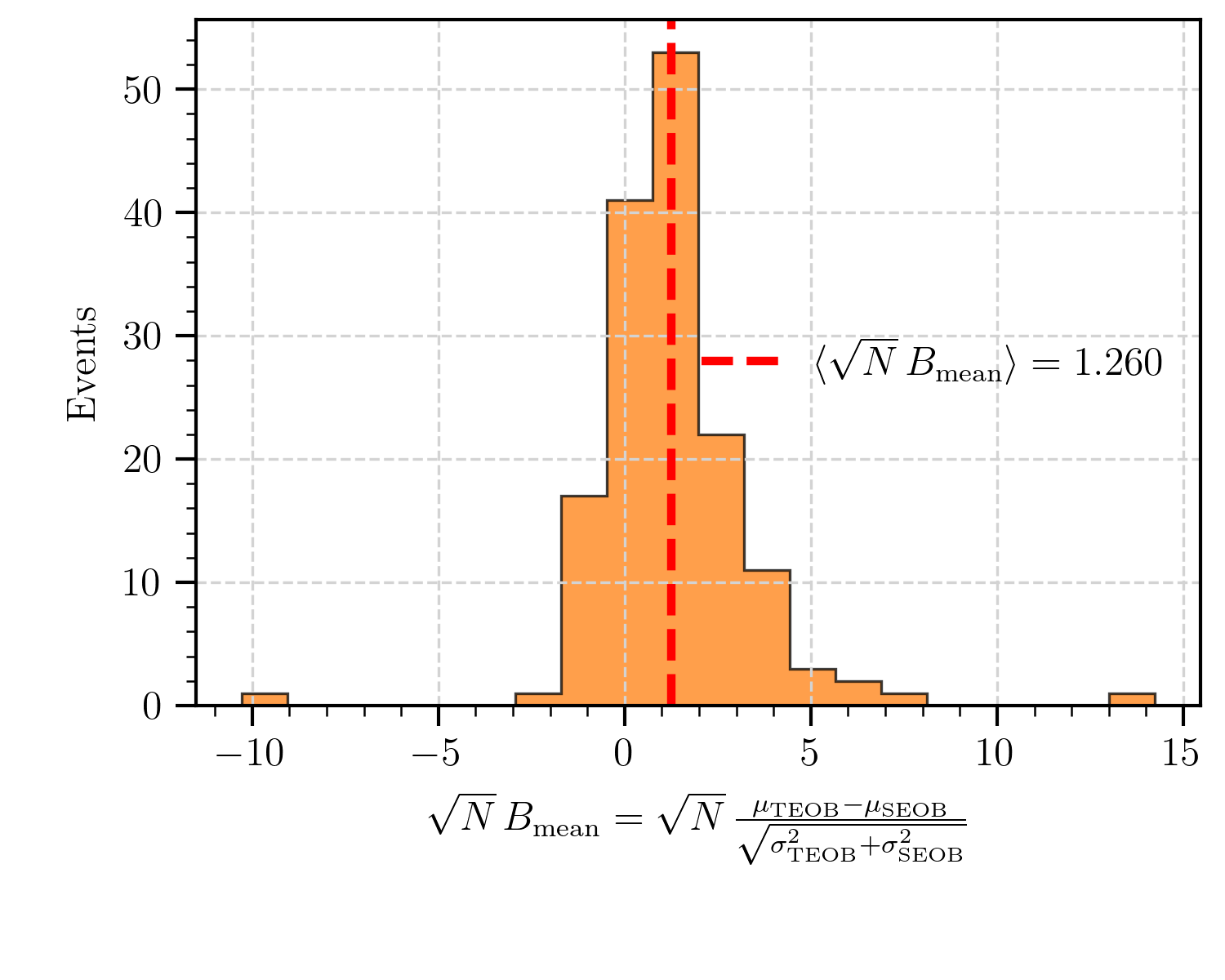}
    \includegraphics[width=0.49\textwidth]{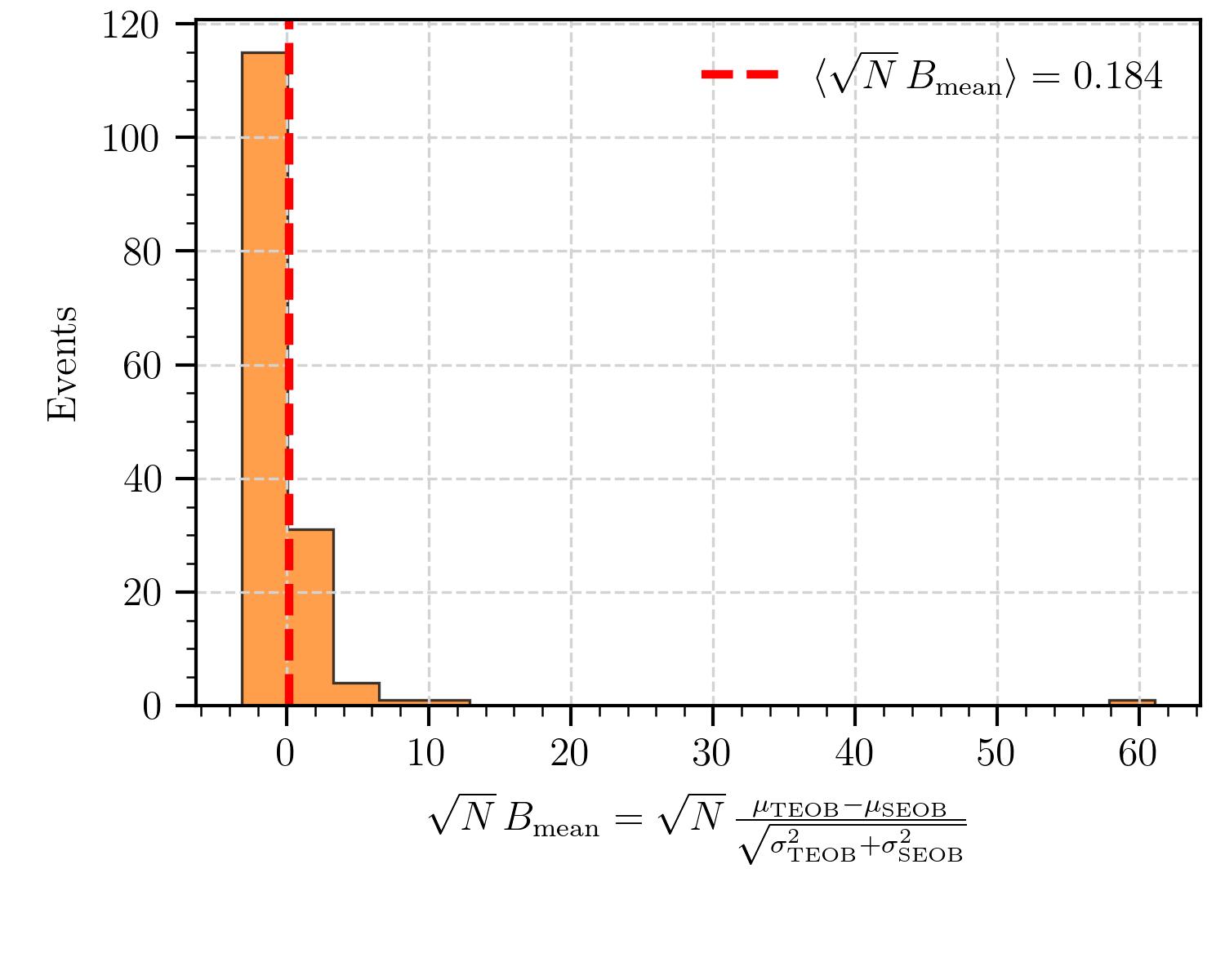}
    \includegraphics[width=0.49\textwidth]{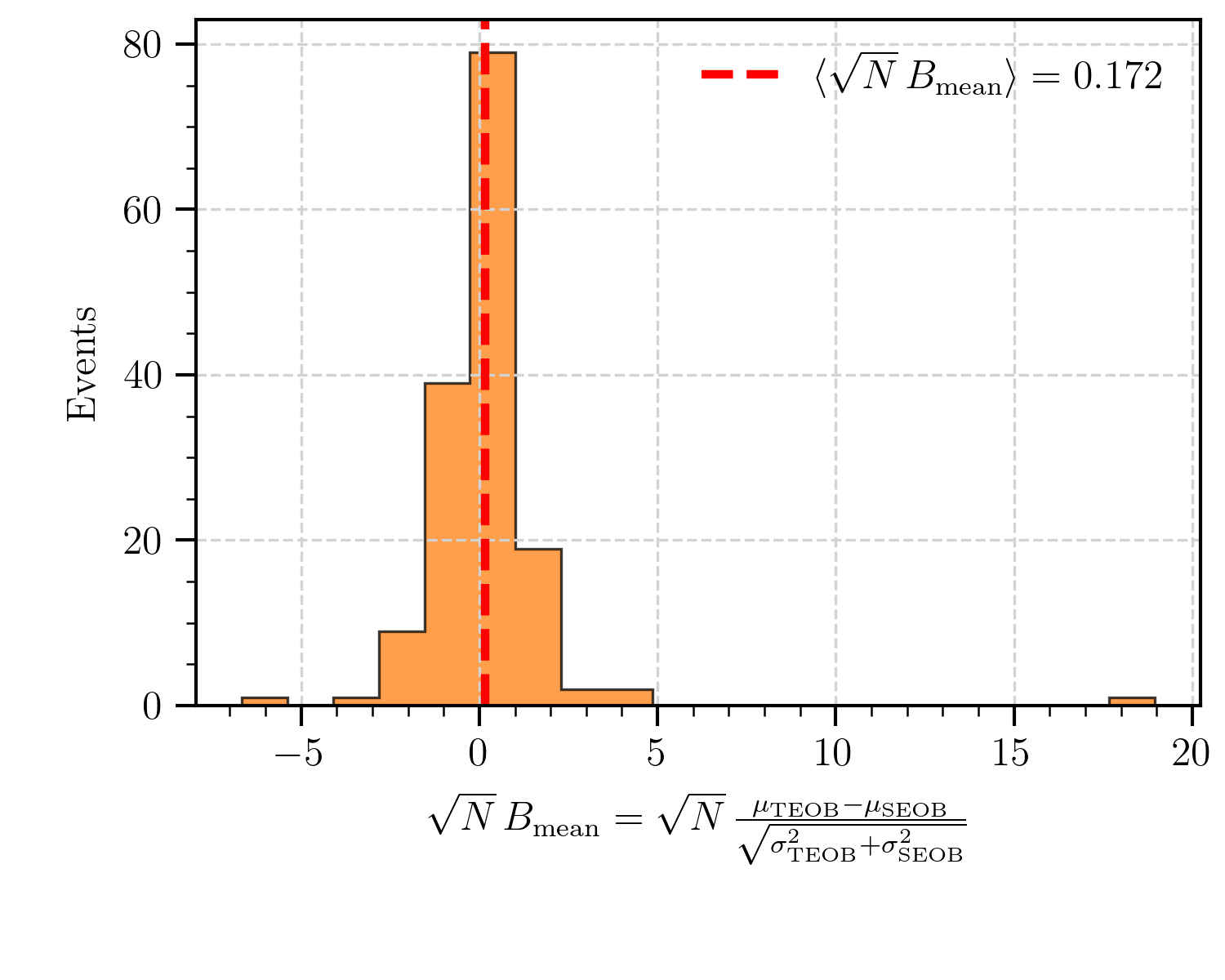}
    \caption{This figure shows the systematic shift of mean scale with $\sqrt{N}$, where $N$ is the number of events. The red dashed line shows the equal mean of all the events. The upper left panel shows the redshift, upper right panel shows the effective spin, lower left panel shows the eccentricity and lower right panel shows the mass ratio. This effect gets significant when it accumulates at population level, it can been seen in the redshift and effective spin plot in top panel.}
    \label{fig:pop_signed_bias}
\end{figure*}

\begin{figure*}
    \centering
    \includegraphics[width=0.48\textwidth]{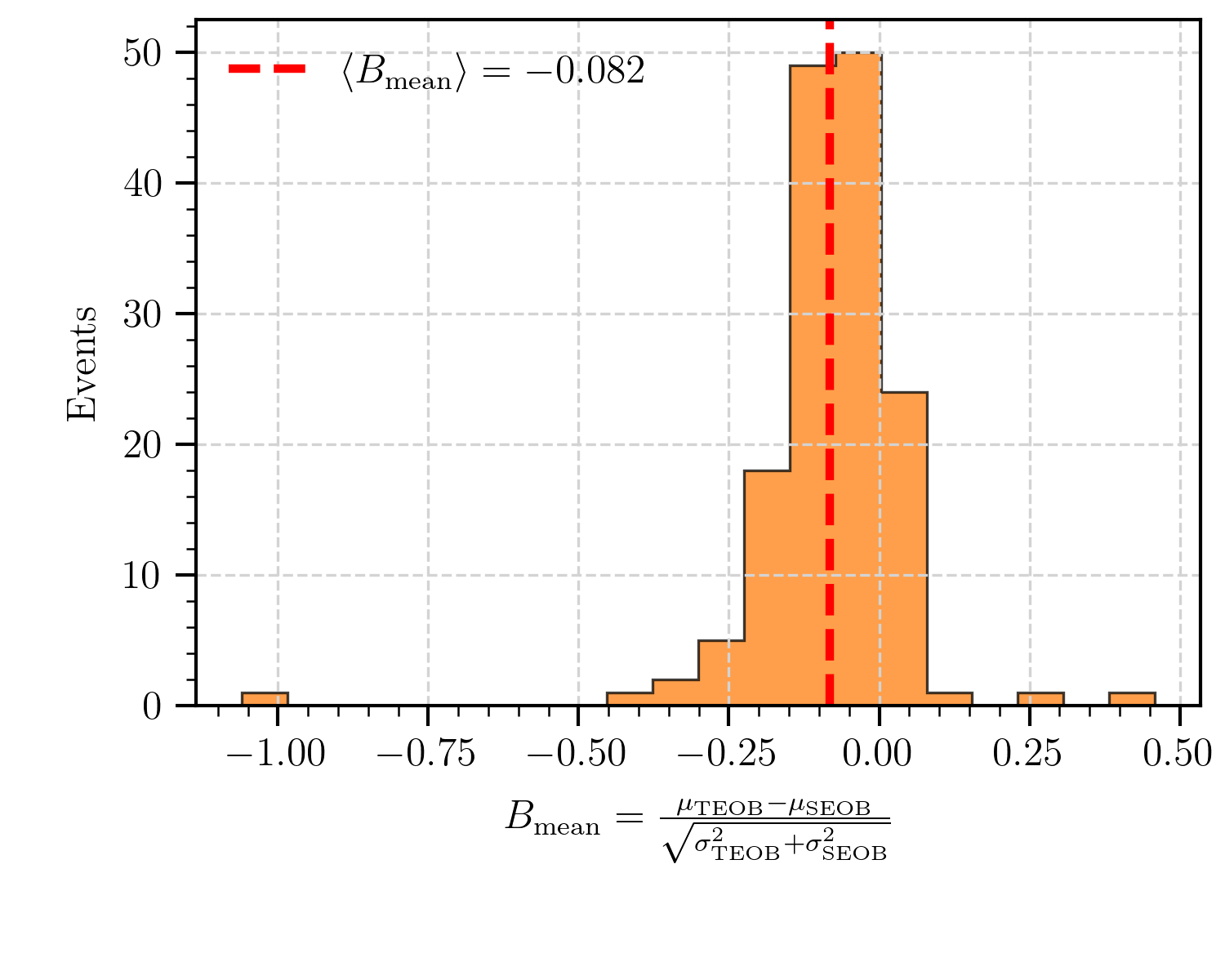}
    \includegraphics[width=0.48\textwidth]{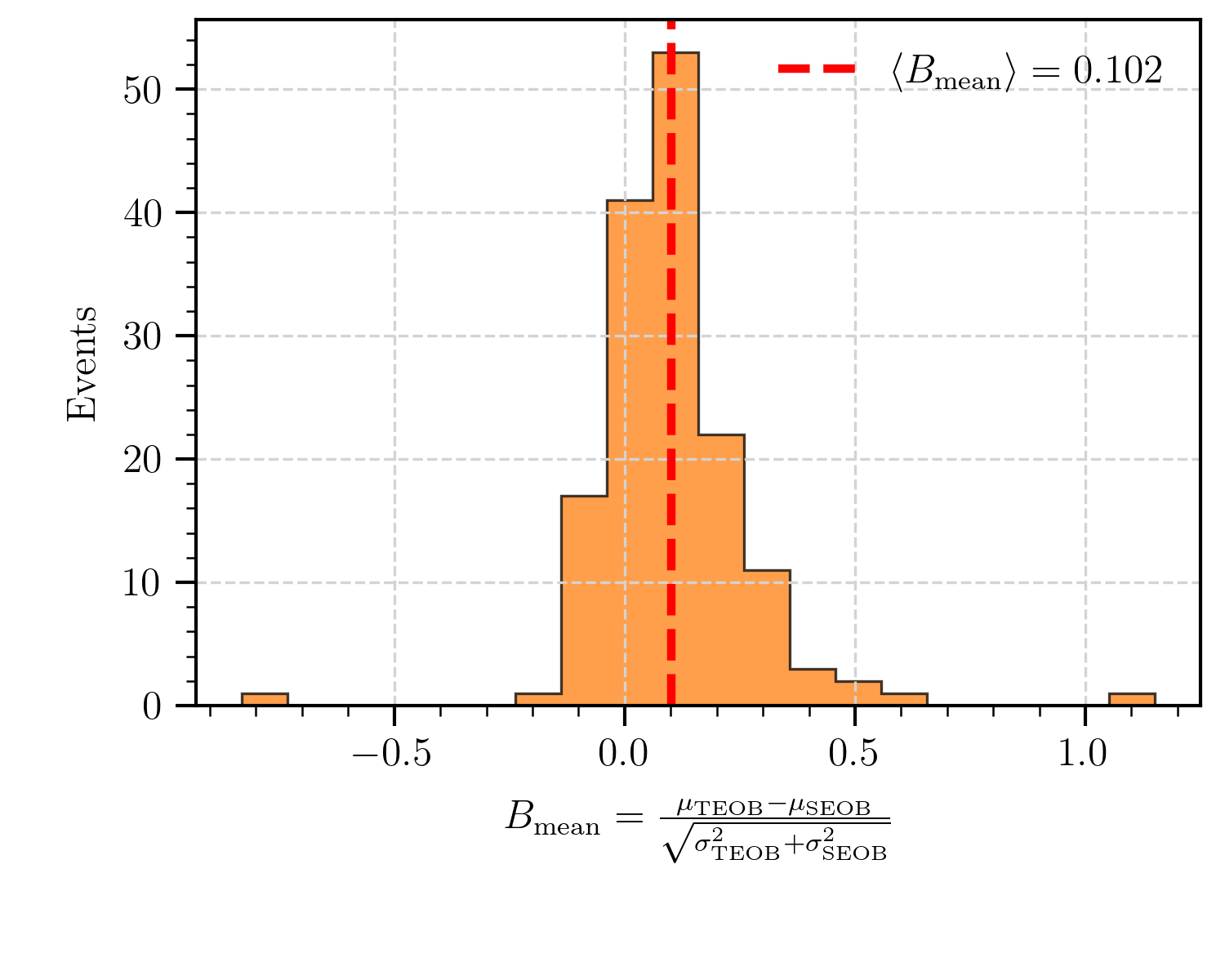}
    \includegraphics[width=0.48\textwidth]{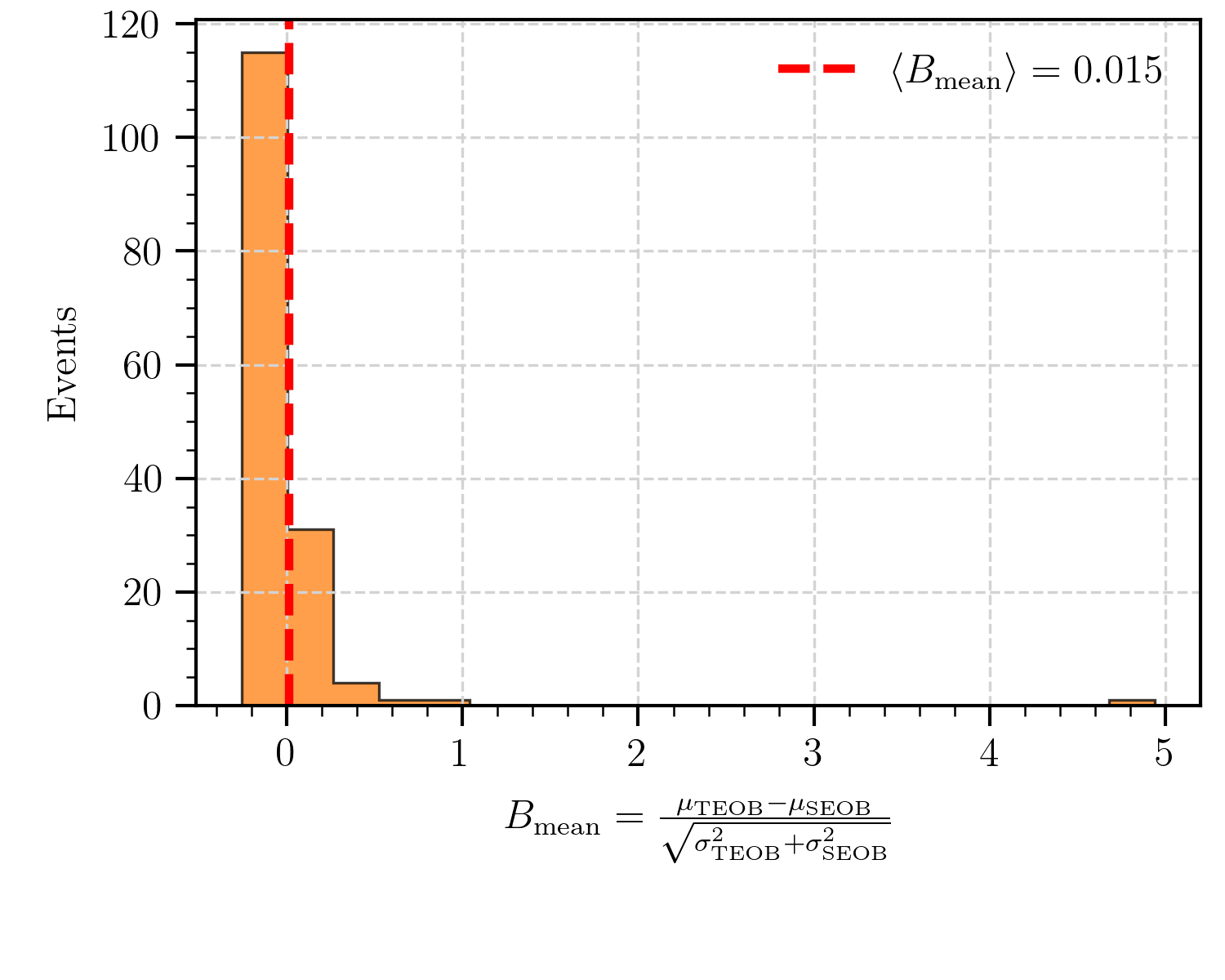}
    \includegraphics[width=0.48\textwidth]{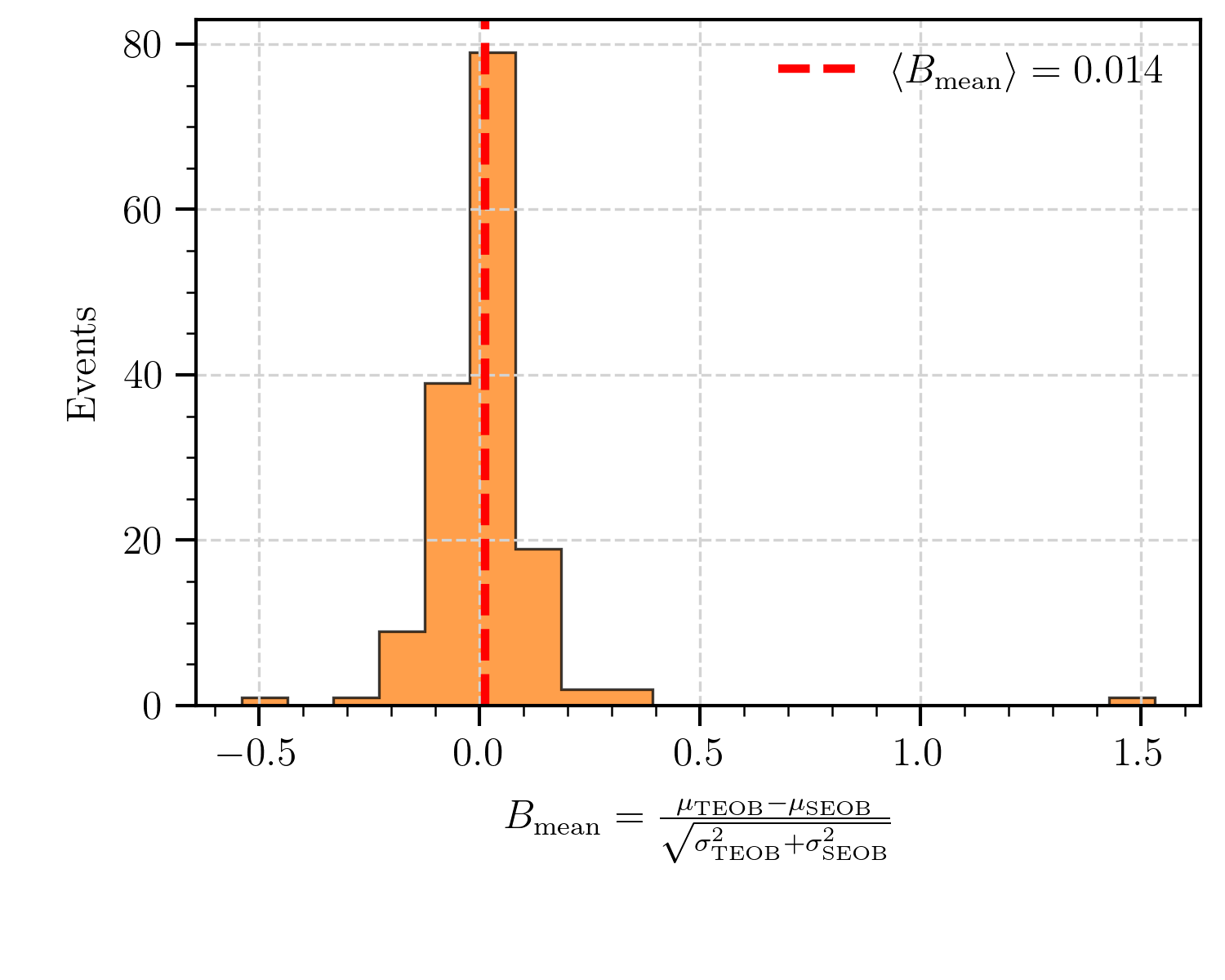}
    \caption{This figure shows the individual event systematic shift of mean value for each event. The red dashed line shows the equal mean of all events. The upper left panel shows the redshift, upper right panel shows the effective spin, lower left panel shows the eccentricity and lower right panel shows the mass ratio. At the event level, both waveform seems to be consistent with each other.}
    \label{fig:ind_mean}
\end{figure*}

We translate  these visible-to-the-eye diagnostics into a few quantities that assess whether these trends exhibit a
bias.  The first and simplest diagnostic uses the classic frequentest test: we compare the sample means obtained using both methods to each other, accounting for the width of each measurement.
Figure~\ref{fig:pop_signed_bias} has four panels, each showing a histogram of the per-sample z-score-like quantity $B_x$:
\begin{align}
B_x =\sqrt{N}
\frac{\mu_{x, TEOB} - \mu_{x, SEOB}
 }{
  \sqrt{\sigma^2_{x,TEOB}+\sigma^2_{x,SEOB}}
  }
\end{align}
where the factor of $\sqrt{N}$ accounts for the additional resolving power afforded by combining events. 
In each panel, the vertical red line show the average of these quantities.  Most are qualitatively consistent with unity.
However, the redshift distribution is much more sensitive to distant sources than to close-in sources, so this simple diagnostic understates the impact of these subtle biases.
On an event-by-event basis, however, these differences are not at all apparent, with shifts in the mean much smaller
than one standard deviation.  To highlight the subtlety of this cumulative bias Figure~\ref{fig:ind_mean} shows the same
data, without the factor of $\sqrt{N}$; in this representation, any two events' posteriors have nearly indistinguishable
means.

\subsection{Waveform systematics propagate into population inference}
\label{subsec:population_level_systematics}

\begin{figure*}
    \centering
    \includegraphics[width=0.49\textwidth]{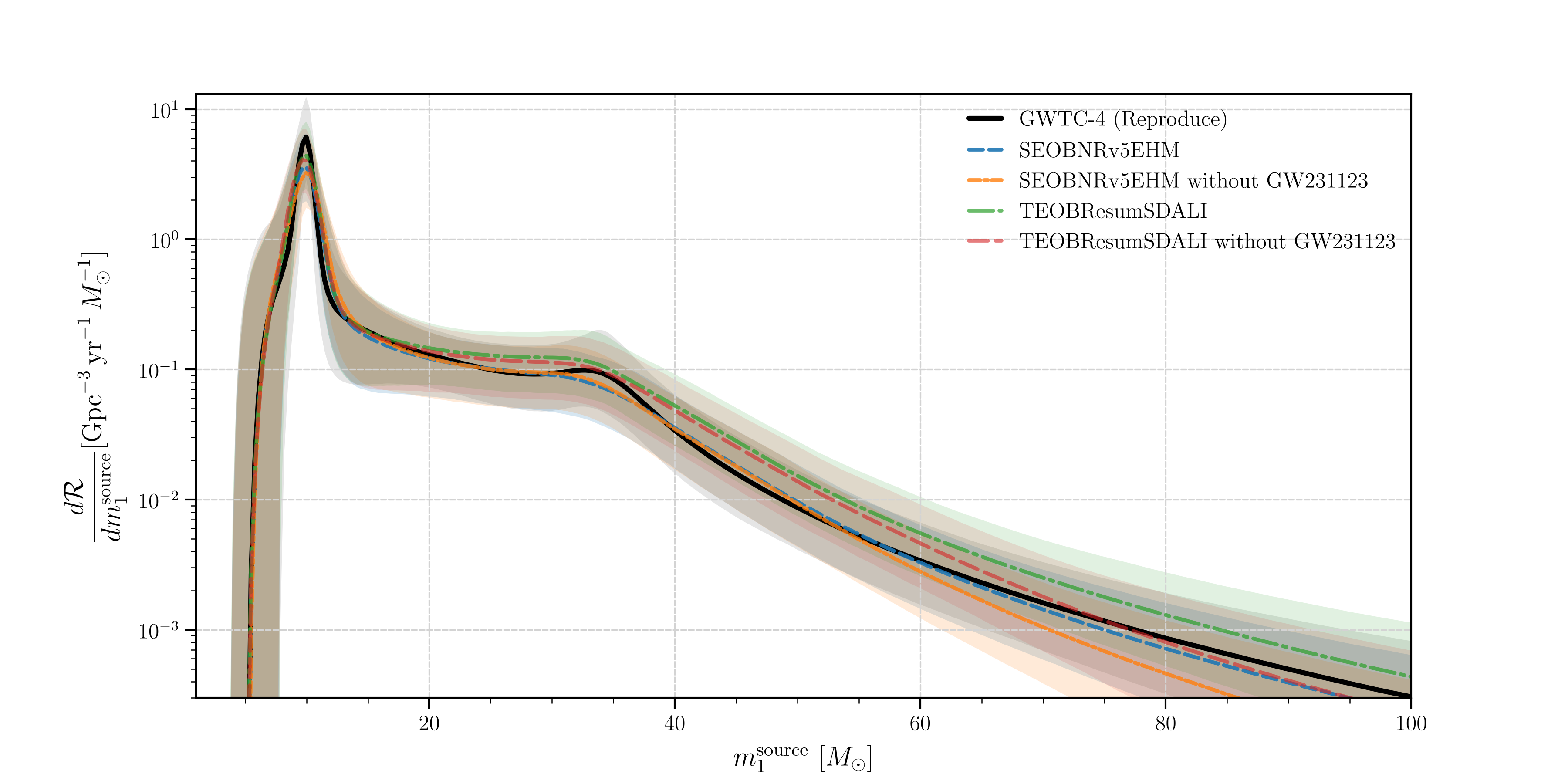}
    \includegraphics[width=0.49\textwidth]{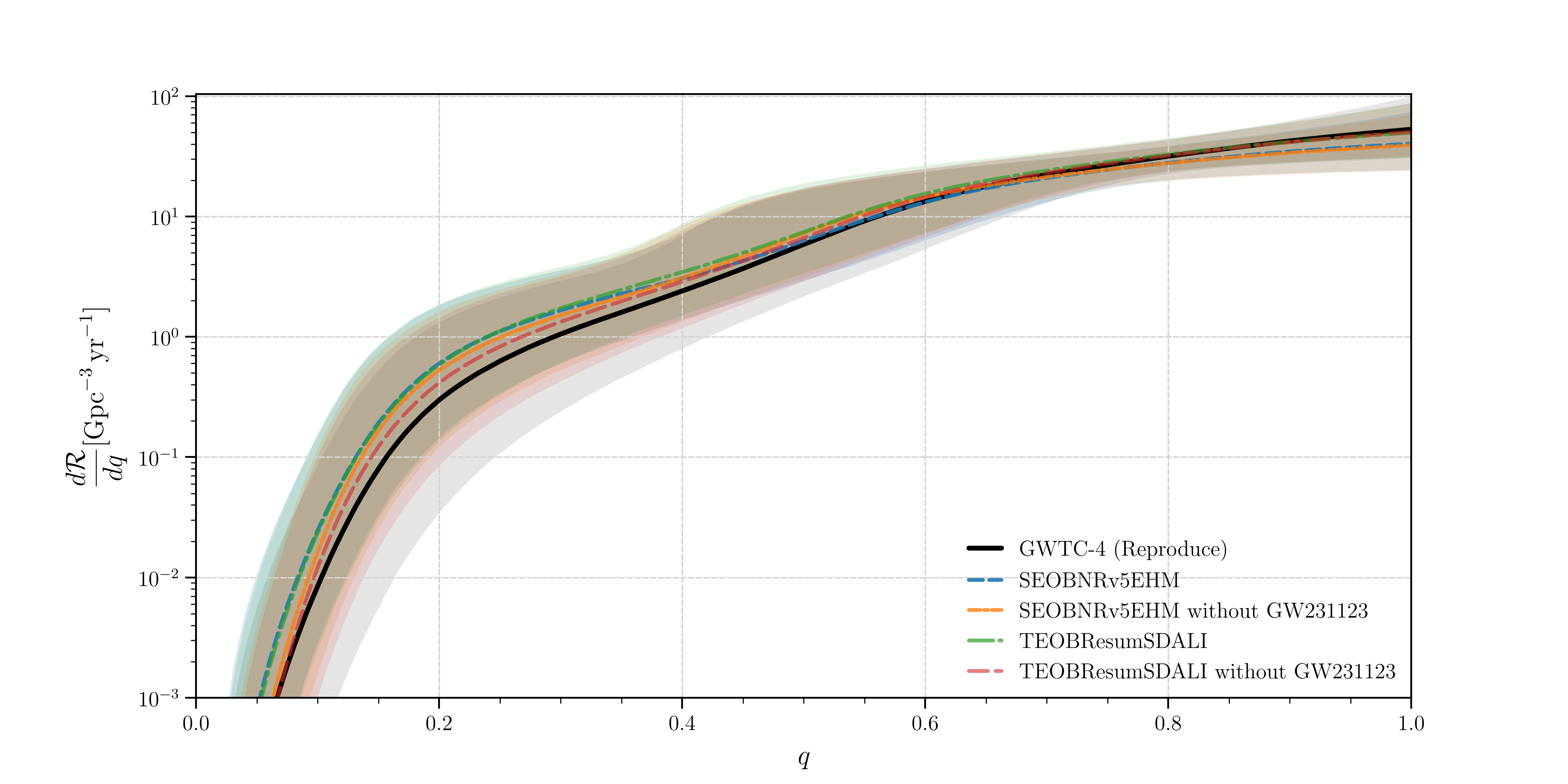}
    \includegraphics[width=0.49\textwidth]{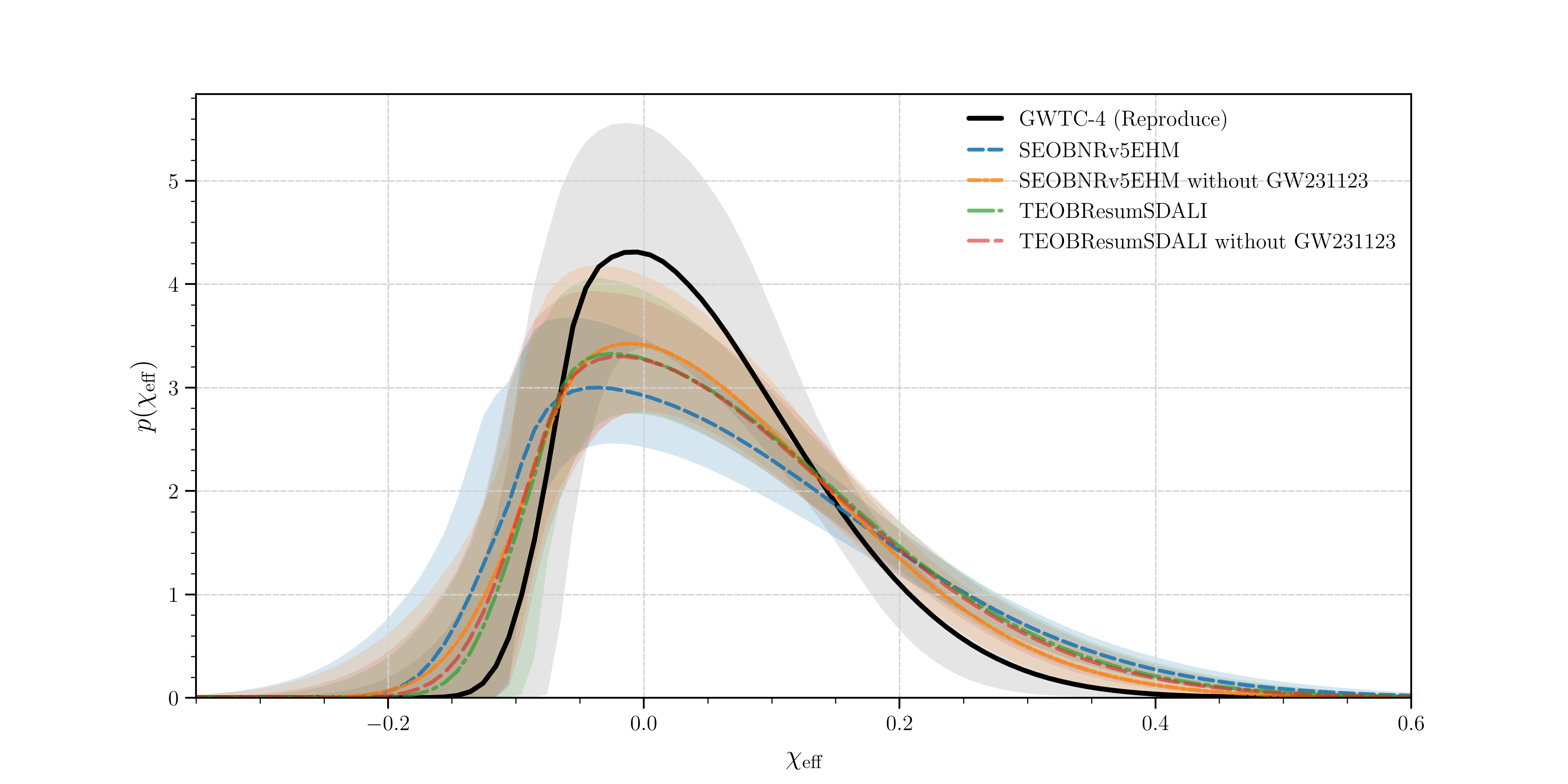}
    \includegraphics[width=0.49\textwidth]{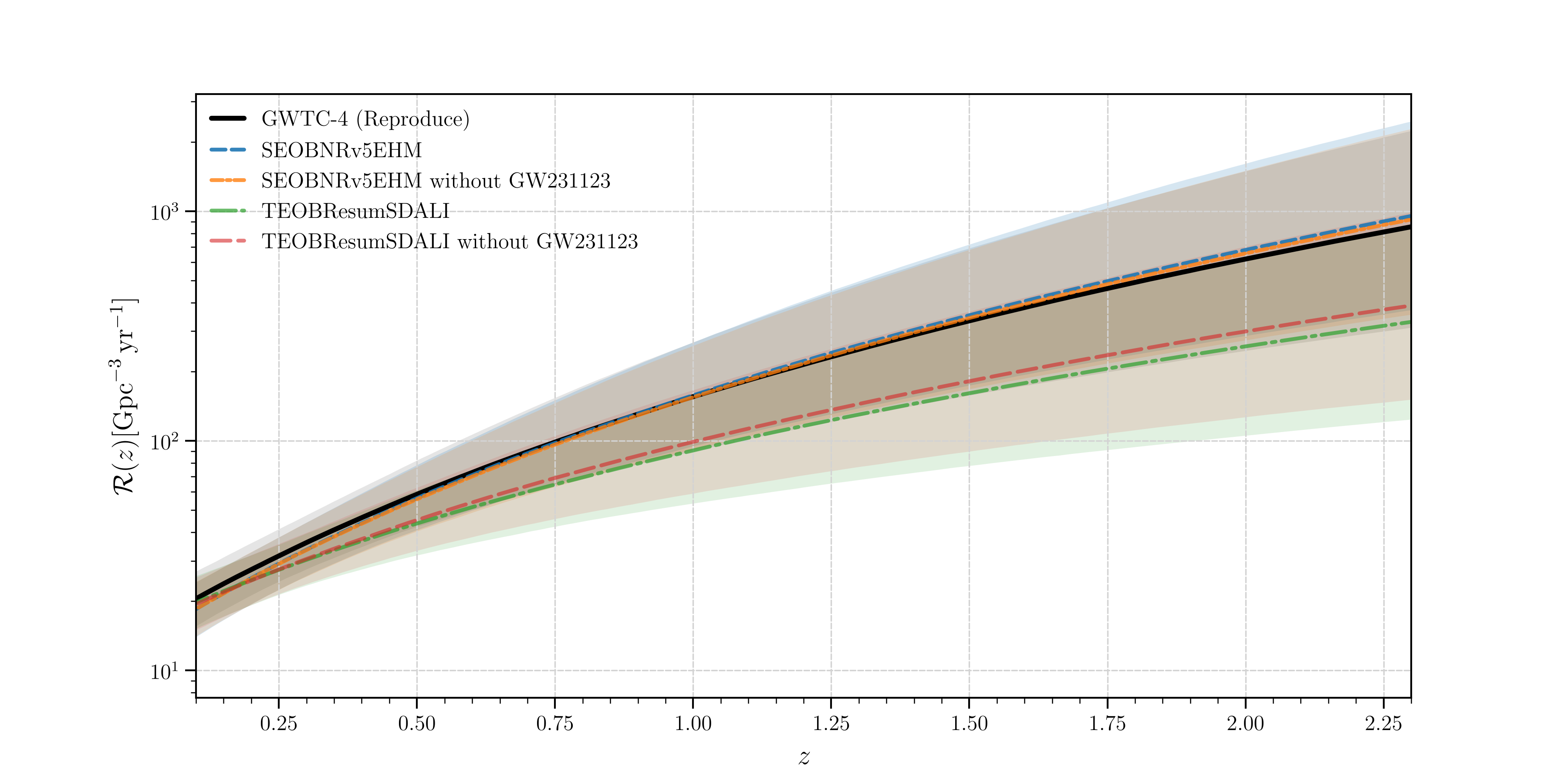}
    \caption{This shows the comparison of GWTC-4 studies assuming quasi-circular binaries vs RIFT PE using \textsc{SEOBNRv5EHM} and \textsc{TEOBResumSDALI}. The parameter estimation studies GW231123\_135430 shows the extreme waveform systematics, therefore, we also show the analysis without GW231123\_135430.}
    \label{fig:gwtc4_vs_rift_all}
\end{figure*}

Figure~\ref{fig:gwtc4_vs_rift_all} compares the results of end-to-end population inference performed with \gwk on
multiple waveform inputs; each panel provides the posterior predictive distribution for one population parameter.  In black, we have reproduced the GWTC-4 analysis with our framework, using quasi-circular waveform inputs from
published LVK results. The solid line shows the median; the shaded region as usual shows the 90\% credible region. 
The colored dashed lines and corresponding shaded regions show results for analyses with the two different models.
Though very similar in many respects, these analyses differ from the quasi-circular results and from each other in three
key ways: the redshift distribution;  the spin distribution; and the primary mass distribution.


The inferred redshift distribution shows the greatest difference between the two waveform models. Our results derived
with \textsc{SEOBNRv5EHM} are fairly consistent with our reanalysis of GWTC-4 with quasi-circular waveforms.  By contrast, the
analysis performed with \textsc{TEOBResumS-Dali} inputs systematically predicts a more gradual evolution of merger rate with redshift,
reflecting the subtle systematic differences in recovered distance shown in Figure~\ref{fig:gwtc4_vs_rift_all}.  To assess
whether these differences reflect disagreement about the handful of rare massive events where the two models are known
to draw different conclusions, this figure also shows analyses that omit GW231123\_135430. We find effectively unchanged results for redshift with this event omitted.

The inferred spin distribution also shows some differences between waveform models: both disagree with one another (and also from the quasi-circular baseline).   However, unlike the redshift distribution, our conclusions are sensitive to the exact waveform catalog used: omitting GW231123\_135430, we find the two models agree with each other (and the TEOBReumS full-catalog result).


Our conclusions about the primary mass and mass ratio distributions are fairly consistent with each other, whether or not massive events like GW231123\_135430 are included. However, with eccentric waveforms $33 M_\odot$ peak is not as sharp as quasi-circular waveform.

\begin{figure}
    \centering
    \includegraphics[width=0.48\textwidth]{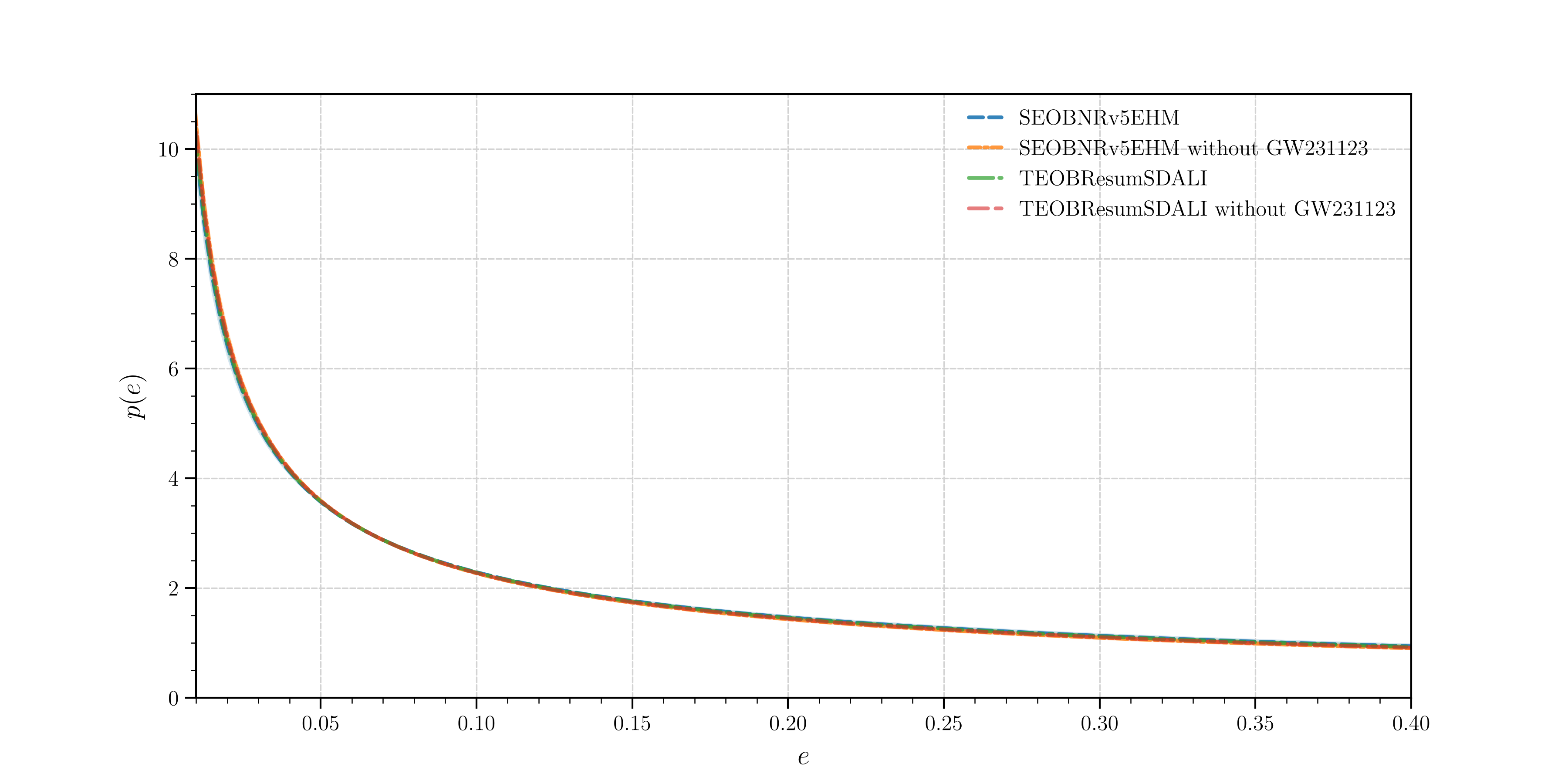} 
    \caption{ This shows the comparison of eccentricity distribution using \textsc{SEOBNRv5EHM} and \textsc{TEOBResumSDALI}. We see the same distribution using the different waveform model, showing almost no bias effect at population level.}
    \label{fig:ecc_dist}
\end{figure}

Figure~\ref{fig:ecc_dist} shows the recovered eccentricity distribution using two waveform models, and they are consistent with each other and we do not observe any significant effect of GW231123\_135430 at the population level or any waveform systematics, regardless of different eccentricity values at parameter estimation level. That effect can also arise due to the dominant circular population.

\subsection{BNS and NSBH population inference}
\label{subsec:bns_nsbh_results}
\begin{figure*}[t!]
    \centering
    \includegraphics[width=0.48\textwidth]{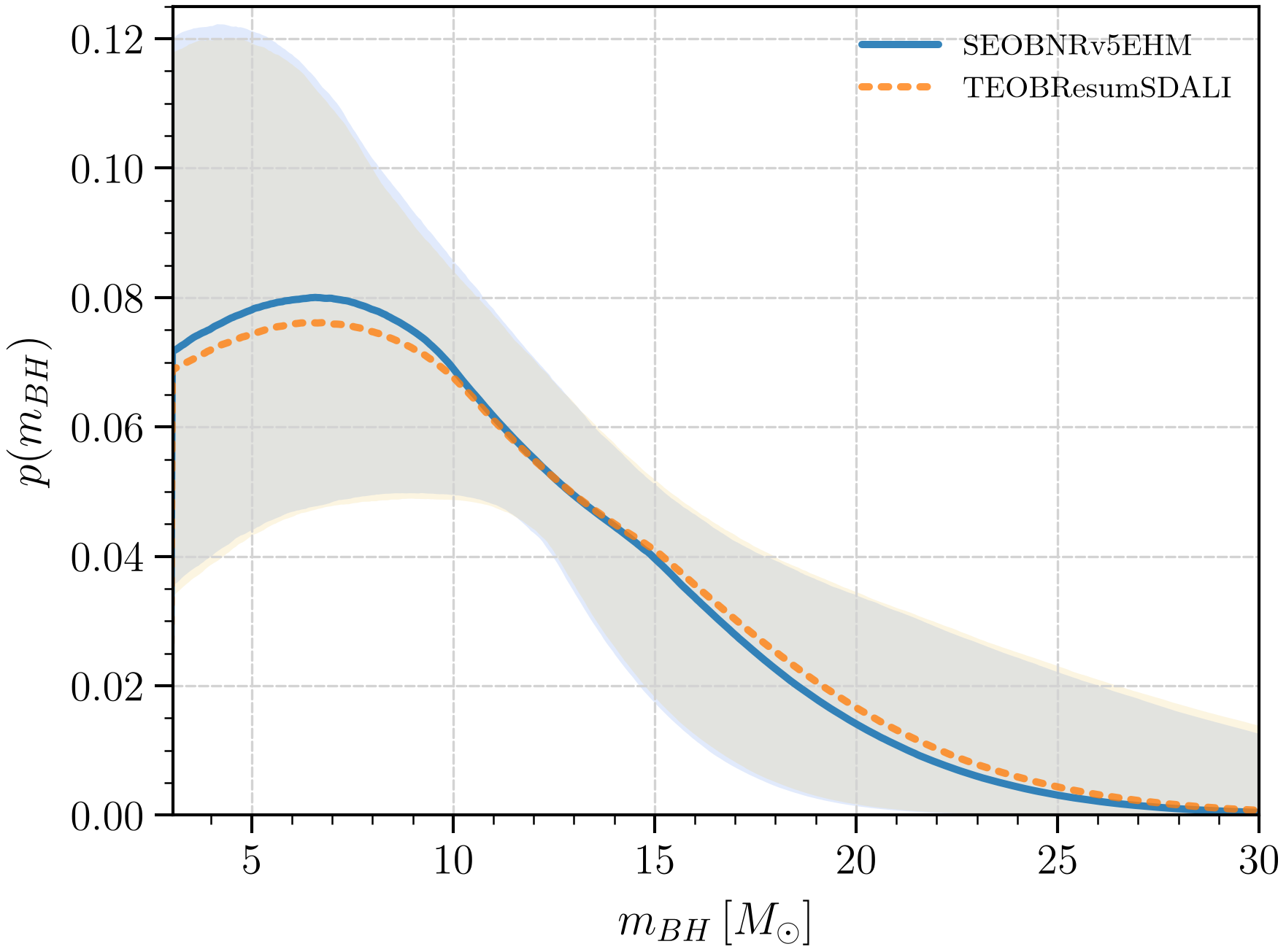}
    \includegraphics[width=0.48\textwidth]{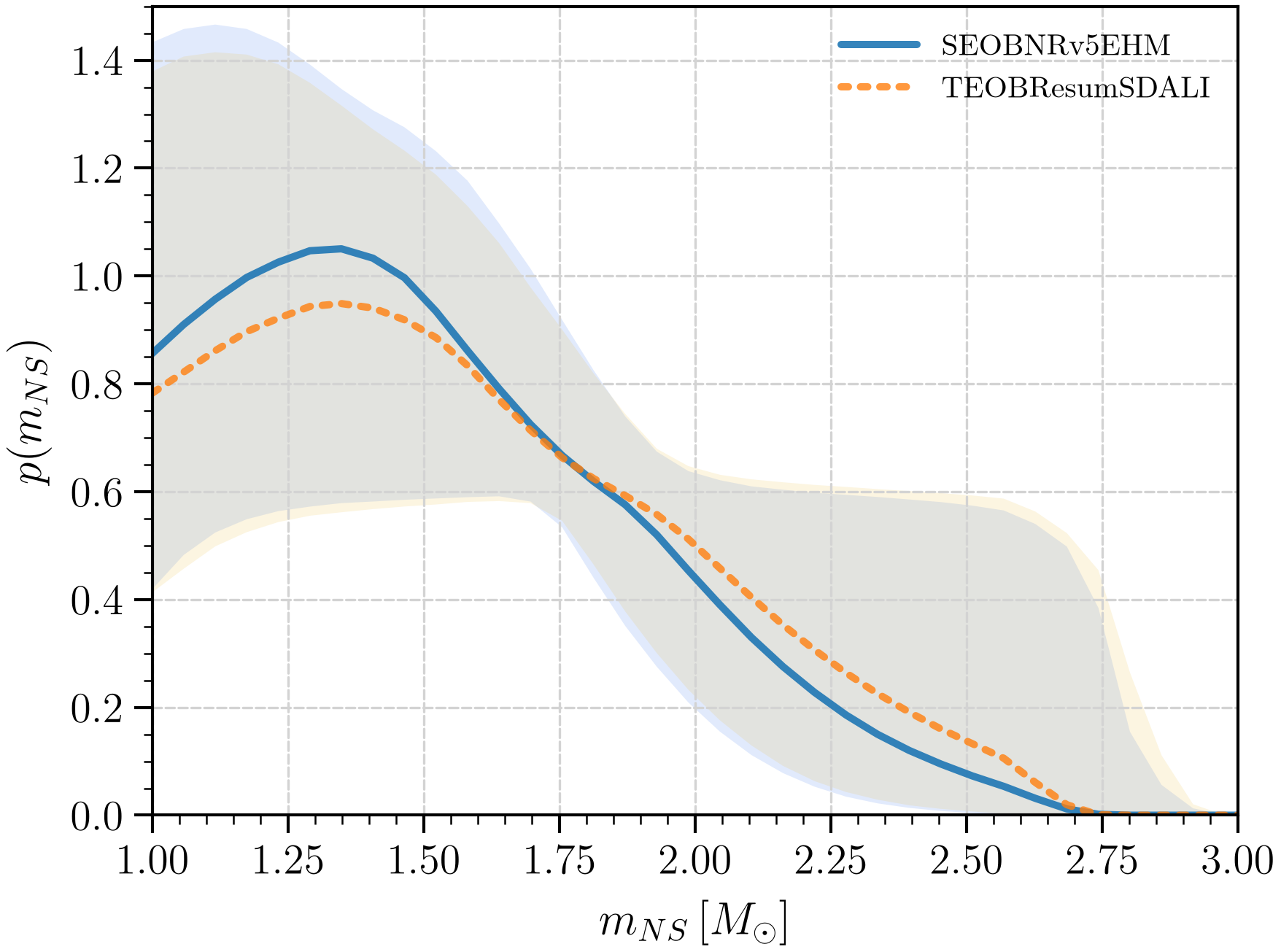}
    \includegraphics[width=0.48\textwidth]{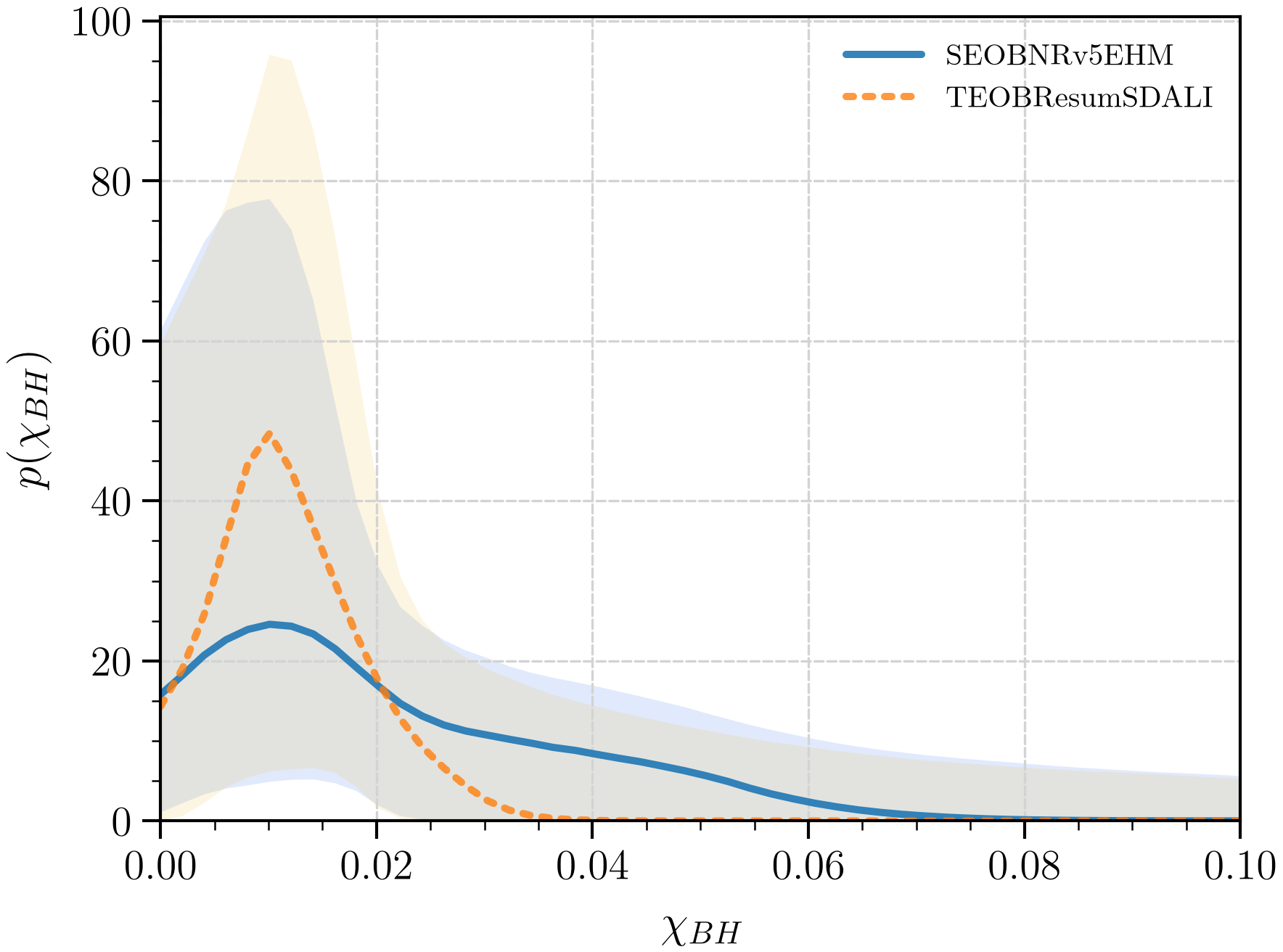}
    \includegraphics[width=0.48\textwidth]{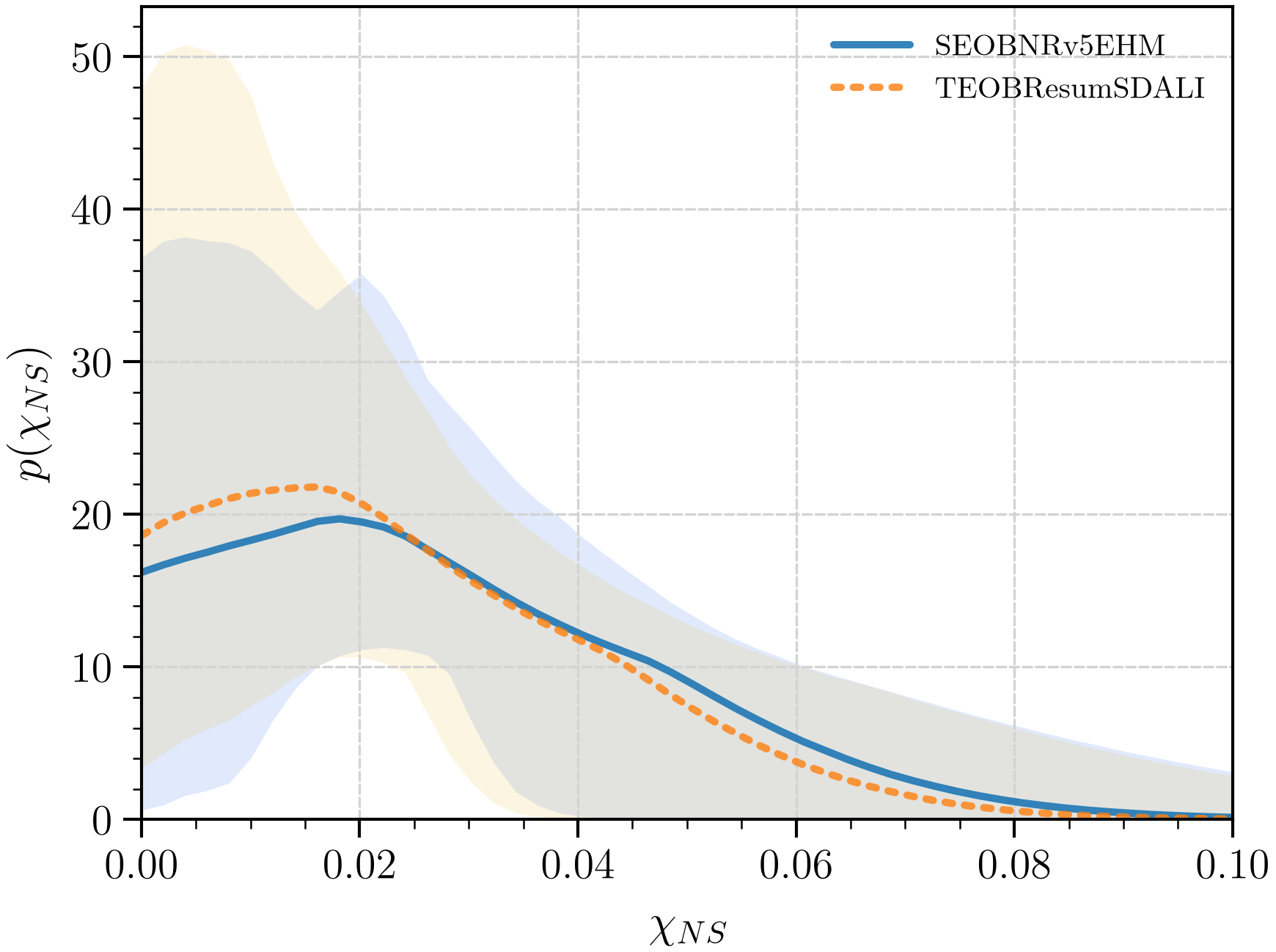}
    \caption{The top left figure shows the mass distribution of BHs in NSBH systems, and the top right figure shows the mass distribution of NSs either in BNS or NSBH. Similarly, lower left figure shows the spin distribution of BHs in NSBH and lower right figure shows the spin distribution of neutron stars.}
    \label{fig:ns_bh_mass}
\end{figure*}

We apply the same population inference framework to the sample of ``NS-BH'' event candidates listed in Table~\ref{tab:bns_nsbh_events}.
Figure~\ref{fig:ns_bh_mass} presents our inferred mass and spin distribution for the BHs (left panels) and NS (right
panels)  in the population of merging NS-BH binaries.  
Our conclusions for the mass and spin distribution for these objects are qualitatively consistent with previously
reported analyses \cite{2023MNRAS.518.5298B,2026arXiv260322461M}, with a few mild differences compared to the first analyses.
First, our BH distribution extends smoothly to $3M_\odot$ -- it exhibits no ``mass gap'' between BH and NS.
Second, as we do not incorporate any nuclear physics modeling or observations into our analysis, the inferred mass distribution for the smaller objects extends
substantially above $2M_\odot$.  The secondary mass distribution's extent is driven by our event list, which includes
GW190814, in contrast to many previous studies of ``NS''-containing binaries in the merging binary population.
Third, we find somewhat less evidence for large BH spins in BH-NS binaries, again informed in part by GW190814. 

\begin{figure*}[t!]
    \centering
    \includegraphics[width=0.48\textwidth]{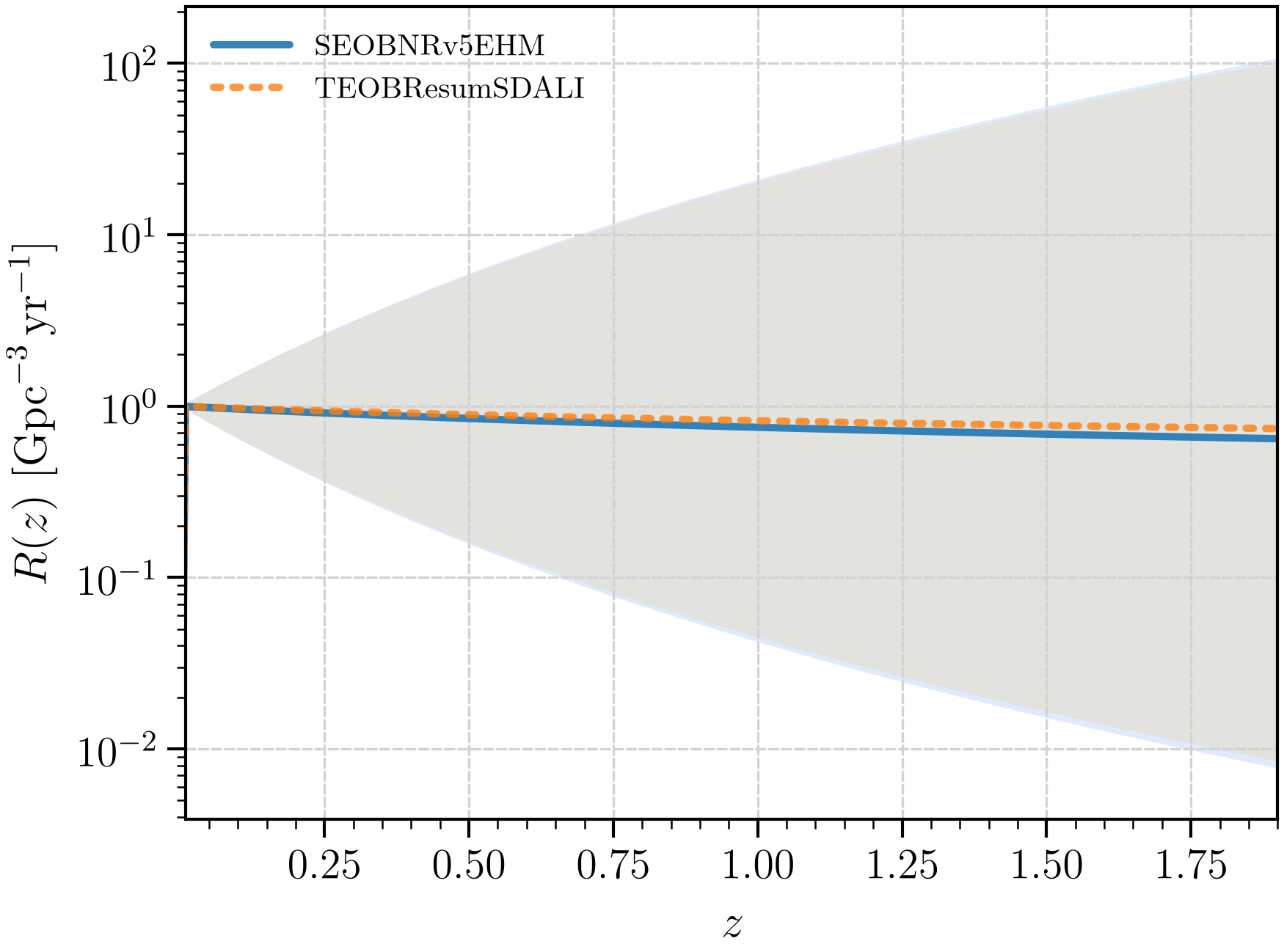}
    \includegraphics[width=0.48\textwidth]{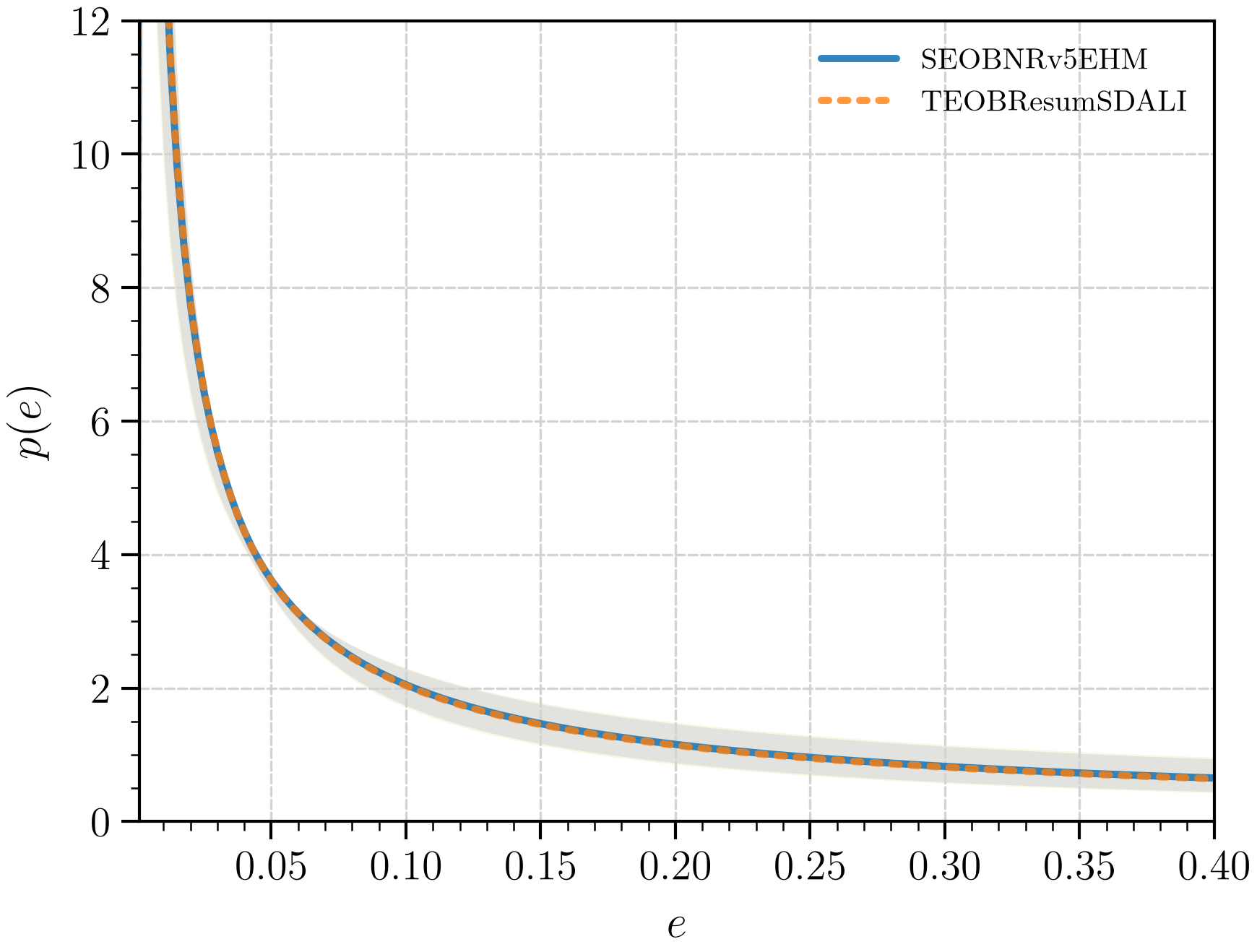}
    \caption{The left figure shows the joint distribution of redshift for BNS and NSBH and right figure shows the joint distribution of eccentricity for BNS and NSBH.}
    \label{fig:ns_bh_redshift_ecc}
\end{figure*}

Figure~\ref{fig:ns_bh_redshift_ecc} shows our inferred redshift and eccentricity distributions.  With so few events over
a narrow chirp mass range, we cannot usefully reconstruct the source redshift distribution with the current sample.
The inferred eccentricity distribution is strongly informed by GW200105\_162426 and our modeling priors, which require a pure
power-law eccentricity distribution for these binaries.
As expected given strong indications of eccentricity from so few event candidates, our recovered population model
predicts that many BH-NS binaries will have substantial eccentricity. At these low chirp masses and high mass ratios,
any signatures of eccentricity should be much more easily identified than the subtle hints of eccentricity present in
shorter, higher-mass event candidates. 

\subsection{Synthetic data population inference}\label{subsec:syn_pop_infer}

We performed a hierarchical population analysis on a synthetic catalog generated from the population model described in Section~\ref{subsec:pop_model_syn}, using posterior samples produced with the RIFT framework. Although the synthetic population contained $155$ binary black hole systems, only $39$ events satisfied the detection criteria used in our analysis. This reduction arises from differences between the signal-to-noise ratio (SNR) estimates employed in our semi-analytical sensitive-volume calculation given in Section~\ref{subsec:hbi_review} and the SNR values recovered through the full RIFT parameter-estimation pipeline. Future studies should therefore adopt a fully self-consistent framework in which the same detection statistic is used during both injection generation and parameter estimation, enabling a more faithful recovery of the detected population.

\begin{figure*}[t!]
    \centering
    \includegraphics[width=0.9\textwidth]{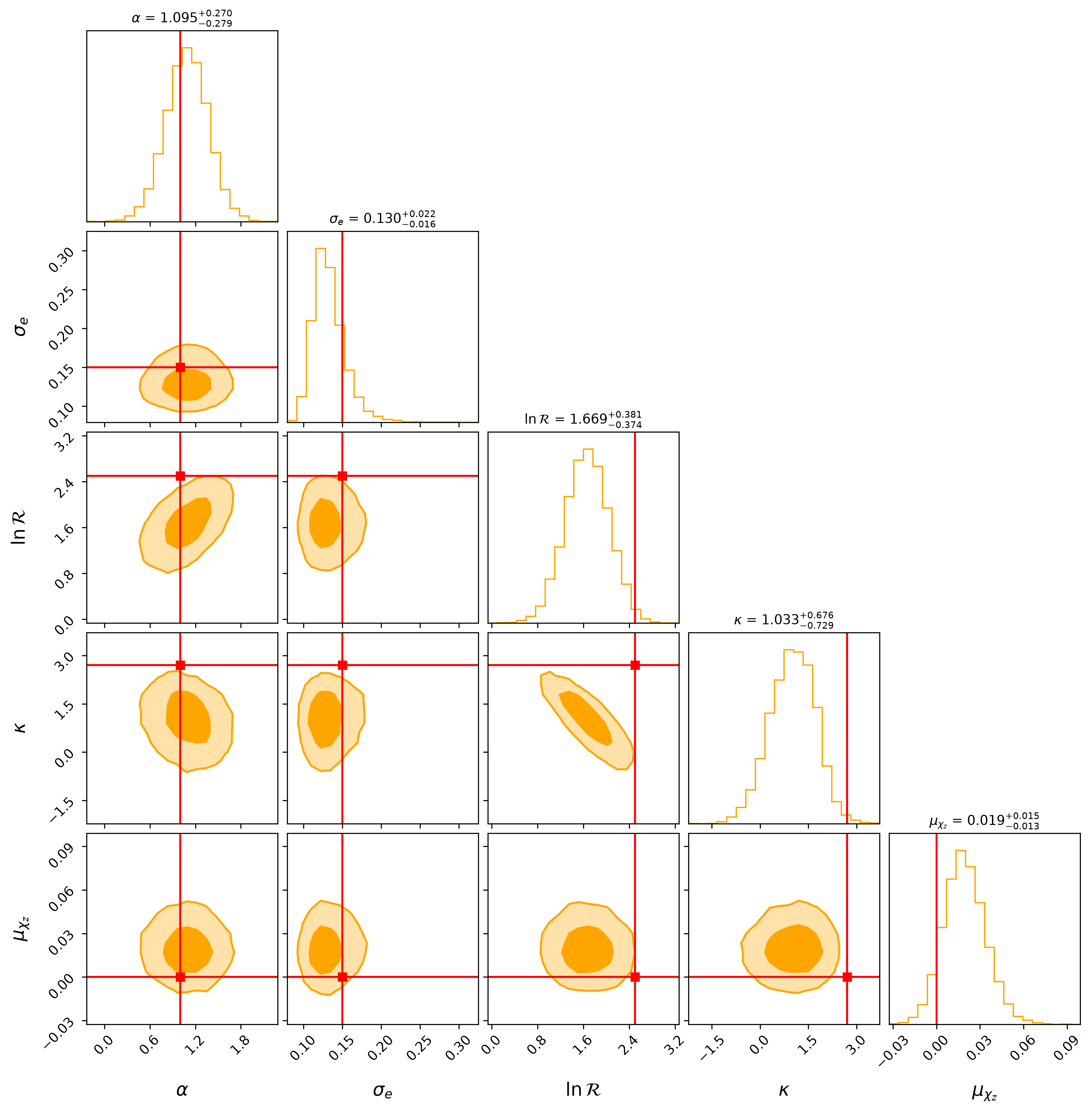} 
    \caption{This figure shows the recovery of hyper-parameter of population model used for synthetic dataset. We were able to use only 39 BBH out of 155 BBH generated with synthetic model due to SNR thresh-hold of detection. However, the results look promising and suggest more precise recovery in future work if we use the full catalog.}
    \label{fig:rift_corner}
\end{figure*}

Despite the limited number of detected events, the recovered hyperparameter posteriors shown in Figure~\ref{fig:rift_corner} demonstrate encouraging agreement with the injected population. In particular, the mass-distribution parameters and eccentricity distribution are recovered reasonably well, while the redshift-evolution parameter $\kappa$ and spin-distribution location parameter remain broadly consistent with their injected values. Although this analysis should be regarded as a preliminary proof of concept due to the limited sample size and simplified selection treatment, these results suggest that the underlying population properties can be recovered reliably with a larger and more self-consistent synthetic catalog.

\section{Conclusion}
\label{sec:conclude}

In this paper, we perform end-to-end population inference on real GW sources using source parameter inferences obtained
with two different waveform models.  We find that the inferred redshift distributions that are conditioned on a specific
waveform model are not consistent with each
other.  We trace the discrepancy between these waveform-specific population inferences to very small but consistent
differences in the recovered distance distribution between these two models.
Otherwise, however, the two models agree with one another and with previously published results, particularly on broad
properties of the BH population.
Our result calls into question the often-accepted conclusion that contemporary GW waveform systematics are more than
sufficient to interpret the contemporary GW population; cf. \cite{2025PhRvX..15c1036D}.
Our investigation strongly suggests that subtle systematic differences can and will build up, impacting conclusions that
rely on multiple events, even in the present-day era.
We recommend further population inferences employ self-consistent inference with multiple waveforms, rather than a
single analysis with mixed samples. 

We also report on joint population inference of the ``NS-BH'' population, including eccentricity.  Unlike other work, we incorporate GW190814
into our joint analysis of this subsample.   Our results corroborate previous analyses \cite{2023MNRAS.518.5298B}, including investigations
which use eccentric source and population inference \cite{2026arXiv260322461M}.

In previous work, we performed population inference end-to-end with a single waveform. In this work, in addition to
performing population inference with two waveforms, we have also  validated our population inference method end-to-end with synthetic data. 
However, all of our analyses in this study employ nonprecessing waveform models. Particularly given the
high interest in and substantial systematics of precession specifically, future studies should also assess similar
issues for binaries with generic  spins.


\section*{Acknowledgements}
This material is based upon work supported by the NSF's LIGO Laboratory, a major facility fully funded by the National Science Foundation. The authors acknowledge the computational resources provided by the LIGO Laboratory's CIT cluster, which is supported by National Science Foundation Grants PHY-0757058 and PHY0823459. ROS acknowledges support from NSF Grant No. AST-1909534, NSF Grant No. PHY-2012057, and the Simons Foundation.


\bibstyle{unsrt}
\bibliography{references_merged}

\end{document}